\documentclass[3p,review,12pt]{elsarticle}
\usepackage{amssymb}
\usepackage{lineno}
\usepackage{float} % pour en incluant une figure
\usepackage{xcolor}%\textcolor[rG-B]{0.00,0.00,1.00}{}
\usepackage{epstopdf}
\usepackage{amsfonts}
\usepackage{array}
\usepackage{booktabs}
\usepackage{longtable}
\usepackage{amsmath}
\usepackage{pdflscape}
\usepackage{graphicx}
\usepackage{pdfpages}
\usepackage{subcaption}
\usepackage{bbm,bm}
\usepackage{stmaryrd}
\usepackage{color}
\usepackage{booktabs}
\usepackage{yhmath}
\usepackage{flexisym}
\usepackage{adjustbox}
\usepackage{physics}
\usepackage{boldline,multirow}
\graphicspath{{./Figures/}}
\usepackage{tabularx}
\usepackage{upgreek}
\usepackage[linesnumbered,ruled,vlined]{algorithm2e}
\usepackage{booktabs}
\usepackage{tabularx}
\usepackage{array}
\usepackage{ragged2e}

\newcolumntype{Y}{>{\RaggedRight\arraybackslash}X}

\usepackage[colorlinks,bookmarksopen,bookmarksnumbered,citecolor=blue,urlcolor=red]{hyperref}

\journal{}

\makeatletter
\def\ps@pprintTitle{%
  \let\@oddhead\@empty
  \let\@evenhead\@empty
  \def\@oddfoot{\hfil\thepage\hfil}%
  \let\@evenfoot\@oddfoot
}
\makeatother
\usepackage{etoolbox}

\makeatletter
\patchcmd{\pprintMaketitle}
  {\hrule\vskip12pt}{\vskip12pt}{}{}
\patchcmd{\pprintMaketitle}
  {\hrule\vskip12pt}{\vskip12pt}{}{}

\patchcmd{\MaketitleBox}
  {\hrule\vskip12pt}{\vskip12pt}{}{}
\patchcmd{\MaketitleBox}
  {\hrule\vskip12pt}{\vskip12pt}{}{}
\makeatother
\begin{document}

\begin{frontmatter}

\title{Human-guided physics-constrained AI agents construct an auditable model of soil-plug evolution}
% \title{Human-guided physics-constrained multi-agent modeling of soil plug evolution during suction-caisson installation}

\author[1]{Jie Shi}
\author[2]{Yimin Lu \texorpdfstring{\corref{cor}}{}}
\ead{Yimin.Lu@ttu.edu}
\author[1]{Zhongkun Ouyang  \texorpdfstring{\corref{cor2}}{}}
\ead{ouyangzk@sz.tsinghua.edu.cn}

\cortext[cor]{Corresponding author}
\cortext[cor2]{Corresponding author}
\address[1]{Institute of Ocean Engineering, Shenzhen International Graduate School, Tsinghua University, Shenzhen 518055, China}
\address[2]{Department of Civil, Environmental, and Construction Engineering, Texas Tech University, USA}

\begin{abstract}
Engineering predictions require physical mechanisms to be translated
consistently into equations, discretization, code, and validation,
yet errors can propagate despite local checks.
Artificial-intelligence (AI) agents automate scientific tasks, but
coordinating and independently auditing the theory-to-solver process
under physical constraints and human oversight remains unresolved.
We introduce a human-in-the-loop, physics-constrained multi-agent
workflow where human experts define admissible physics and modeling
boundaries, while agents retrieve evidence, derive equations,
implement solvers, and audit the theory-to-code chain.
Applied to soil-plug evolution during suction-caisson installation,
the workflow generated and audited 6 formulations in 2.9 h of agent
execution once physical knowledge and inputs were prepared.
Among these formulations, adding seepage-driven soil void-ratio
evolution to the geometric baseline reduced mean absolute final-heave
error from 58.4\% to 9.0\% across 14 profiles; the selected model
further incorporated near-wall dilation and achieved mean absolute
percentage errors of 12.4\% across 9 final-state cases and 4.2\%
at the endpoints of 5 process histories.
Beyond predictive performance, blinded replay recovered all 9 target
problems, while an independent audit uncovered 5 implementation
problems after 36 predefined checks had passed.
Overall, this work extends multi-agent AI beyond task automation
toward human-governed engineering solvers.

\end{abstract}

% No keyword in NC
%\begin{keyword}
%keyword 1 \sep keyword 2
%\end{keyword}

\end{frontmatter}

%=============================================================================
\newpage

%\linenumbers

\section*{Introduction}

\begin{figure}[h!]
\centering
\includegraphics[width=0.85\textwidth]{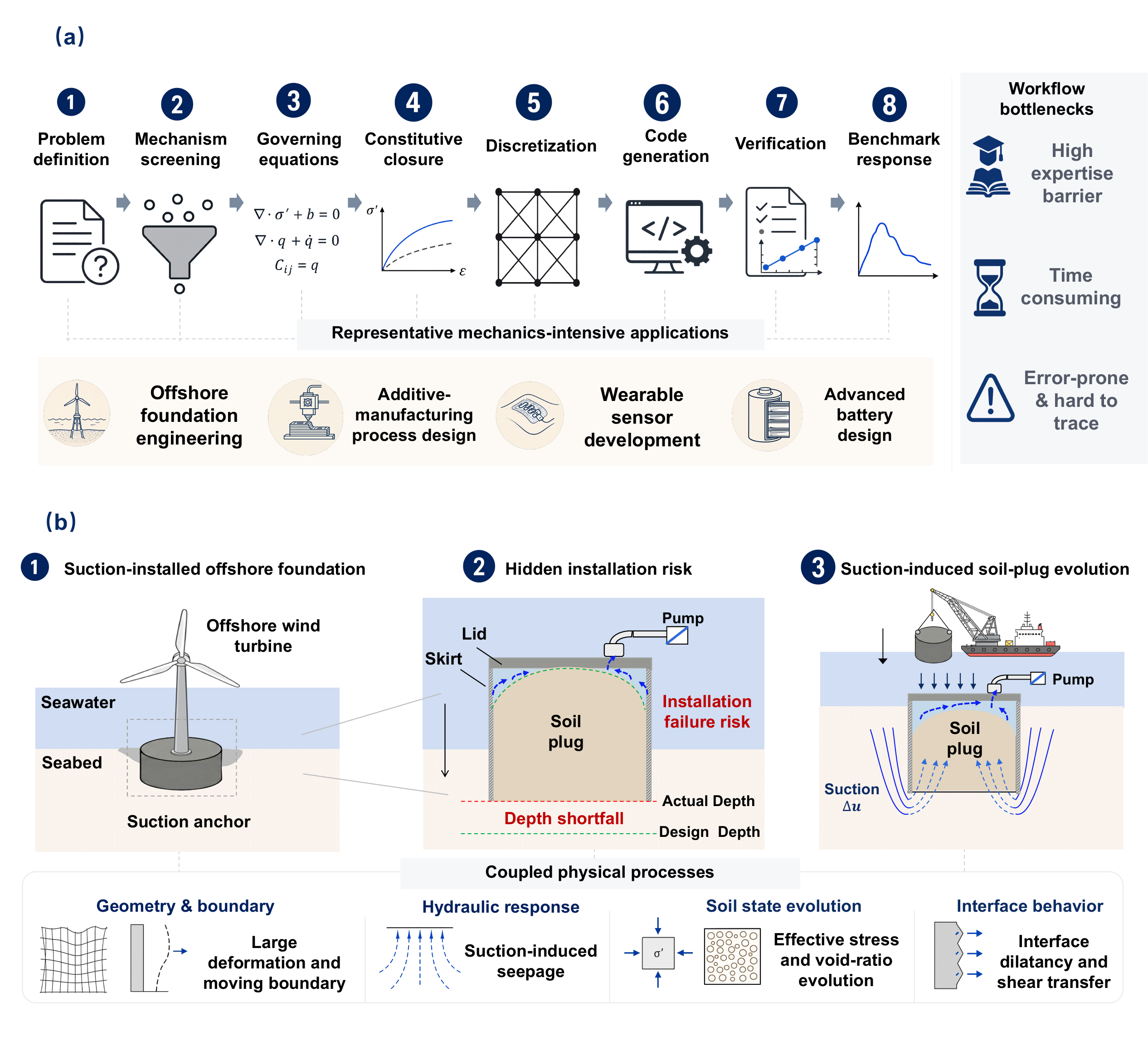}
\caption{The theory-to-solver bottleneck and its demonstration problem. (a) Model development in physics-intensive engineering traverses interdependent stages from problem definition and mechanism selection to governing equations, constitutive closure, discretization, code implementation, verification, and benchmark evaluation. Representative applications illustrate the associated expertise, time, and error-traceability challenges. (b) Soil-plug evolution during suction-caisson installation in sand provides the demonstration problem. The hidden response couples large deformation and a moving boundary, suction-induced seepage, effective-stress and void-ratio evolution, and soil–caisson interface behavior.}
\label{fig:fig1}
\end{figure}

Engineering models become actionable when physical understanding can be translated into executable and validated predictions. This need is especially acute for offshore renewable-energy infrastructure, where rapid deployment must be matched by reliable foundation design under uncertain construction and loading conditions\cite{iea2019offshorewind,gonzalez2024offshorewindwave,zhao2025extremewinds,wu2019foundations,owa2019suction}. Current practice often requires a compromise between case-specific, resource-intensive experiments or high-fidelity simulations and more efficient, generalizable analytical models \cite{owa2019suction,houlsby2005design,harireche2021full}. Developing such models is not a single reasoning task, but a sequence of interdependent decisions spanning problem formulation, mechanism selection, governing-equation derivation, constitutive closure, discretization, code implementation, verification, and benchmark evaluation (Figure~\ref{fig:fig1}(a)) \cite{zienkiewicz1999computational,gens2010soil,wilson2014best}. Each stage draws on distinct expertise, and subtle inconsistencies in assumptions, notation, boundary conditions, or state updates can propagate through the workflow and remain concealed until validation. This long and fragile theory-to-solver process is not unique to offshore engineering; it also arises in physics-intensive problems such as additive-manufacturing process design, wearable-sensor development, and advanced battery design \cite{ates2022wearable,hu2024ampsp,song2024thermalRunaway}.

Suction caissons make this general difficulty concrete. These foundations use a pressure difference across the caisson lid to drive penetration into the seabed, offering installation advantages in energy demand, noise, and soil disturbance \cite{owa2019suction,houlsby2005design}. However, during installation in sand, soil entering the caisson can rise above the surrounding seabed and form a hidden internal plug. Its continued growth may impede penetration, leave the caisson short of its design depth, and introduce uncertainty into foundation performance \cite{owa2019suction,tran2008variation,ragni2020observations,kim2020soil,shi2026installation}. Predicting this evolution requires the simultaneous treatment of a moving boundary, suction-induced seepage, changes in effective stress and soil state, and shear and dilatancy along the soil–caisson interface (Figure~\ref{fig:fig1}(b)) \cite{zhang2024microscopic,mu2026experimental}. Since these processes occur inside a buried and evolving domain, experimental observations are incomplete, high-fidelity simulations are often case-specific, and reduced-order theory is difficult to formulate and implement consistently. These challenges raise the following question: can we establish an efficient, generalizable, physics-constrained, and auditable theory-to-solver framework by leveraging existing knowledge and emerging artificial-intelligence (AI) techniques?
 
Large language models (LLMs) and LLM-driven multi-agent systems offer new capabilities for coordinating such workflows through natural-language reasoning, tool use, information retrieval, and conversion of modeling logic into executable operations \cite{yao2023react,guo2025deepseek,boiko2023autonomous}. Retrieval-augmented generation (RAG) can ground model outputs in external sources \cite{lewis2020retrieval}, while skill-routed retrieval can direct the search to task-specific procedures or knowledge modules \cite{wei2026skill}. However, retrieval relevance is not equivalent to physical admissibility. Before a relation can enter a mechanics model, its provenance, variables, units, assumptions, boundary conditions, and range of validity must be explicit, and it must remain compatible with the other model components. Even a relation drawn from a credible source can yield an invalid solver if it is applied outside its admissible conditions or connected incorrectly to surrounding equations.

Theory implementation and code generation pose a related but distinct challenge. Recent studies have demonstrated the potential of LLMs to translate mathematical formulations into scientific-computing code, including partial-differential-equation (PDE) solvers and geotechnical finite-element models (FEMs) \cite{li2026codepde,kim2025chatgpt}. Multi-agent architectures further decompose scientific workflows by assigning derivation, implementation, and verification to specialized roles \cite{bran2024augmenting,park2026self}. However, this decomposition also introduces handoffs among physical assumptions, continuous equations, discrete updates, and code states. Errors at these interfaces can produce software that is syntactically correct and may even pass local tests while remaining inconsistent with the intended physics or obscuring the path from source evidence to prediction \cite{stechly2025self,dolcetti2025dual,tihanyi2025secure}. Therefore, the central methodological challenge is how knowledge retrieval, physical reasoning, numerical implementation, and validation can be coordinated under a common set of constraints and made auditable across the complete theory-to-solver chain.

In this study, we introduce a human-in-the-loop, physics-constrained multi-agent workflow for constructing, implementing, and validating engineering models across the complete theory-to-solver sequence. Human experts define the modeling objectives, admissible physical knowledge, parameter-source rules, and operational boundaries, while specialized agents collaborate on knowledge retrieval, mechanism formulation, equation derivation, numerical implementation, and independent audit. We demonstrate the workflow by developing a soil-plug evolution model for suction-caisson installation that unifies geometric displacement, seepage-induced soil-state evolution, and soil–caisson interfacial dilatancy through solid-volume conservation. The model-development process is examined through retained workflow records and role-separated audits, while the resulting solver is evaluated using staged benchmark evidence that covers formula reduction and volume closure, final state prediction, and continuous process prediction. The results show that the workflow can produce an accurate and traceable engineering solver while identifying cross-layer inconsistencies that conventional implementation checks may overlook, providing a general framework for auditable multi-agent modeling in physics-intensive engineering.

%=============================================================================
\section*{Results}
\label{results}

%=============================================================================
\subsection*{Workflow-guided model evolution yields the selected soil-plug formulation}
\label{soil-plug-evolution-model}

\begin{figure}[!t]
\centering
\includegraphics[width=0.85\textwidth]{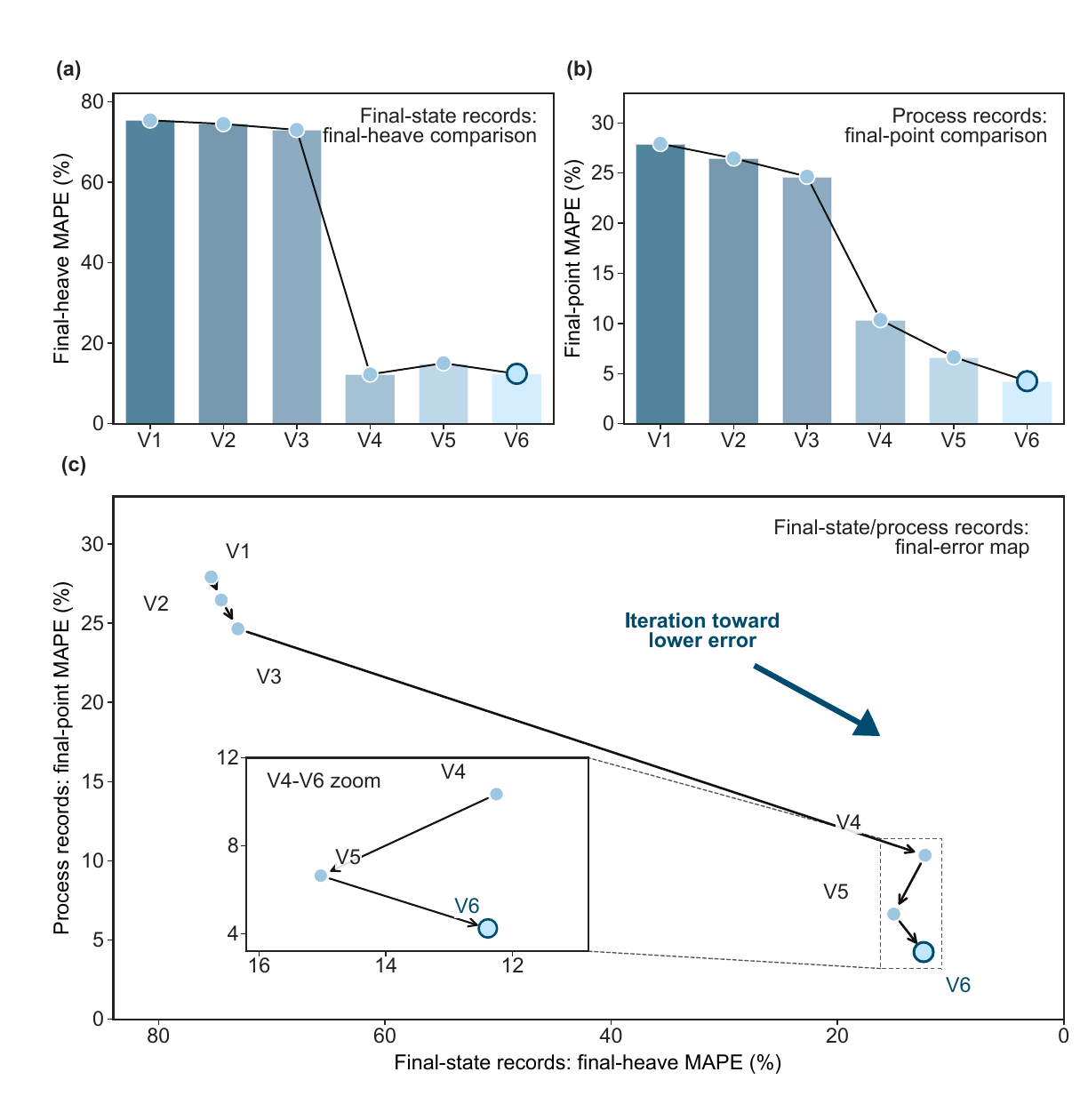}
\caption{Historical screening of candidate soil-plug formulations. All 6 formulations were evaluated using the same benchmark cases and parameter-source rules. 
(a) Final-heave mean
absolute percentage error (MAPE) across 9 final-state cases. (b) Process final-point
MAPE across 5 measured heave--depth histories. (c) Joint evolution of the
two errors across successive formulations; the inset enlarges the differences among V4--V6. Lower values indicate closer agreement with the benchmark observations.}
\label{fig:candidate-screening}
\end{figure}

We applied the physics-constrained multi-agent workflow to develop 6 candidate formulations
(V1--V6) for soil-plug evolution during suction-caisson installation. Each version arose from a diagnostic question prompted by the performance of the preceding formulation and was evaluated using the same historical benchmark cases and parameter-source rules. Screening based on final-heave and process final-point errors identified V6 as the selected formulation since it retained a low final-heave error while producing the lowest process final-point error (Table~\ref{tab:soil-plug-version-history} and Fig.~\ref{fig:candidate-screening}). Model development proceeded through three diagnostic cycles.

\begin{table}[h!]
\centering
\scriptsize
\setlength{\tabcolsep}{3.0pt}
\renewcommand{\arraystretch}{1.02}
\caption{Workflow-guided development and historical screening of candidate soil-plug
formulations.}
\label{tab:soil-plug-version-history}
\begin{tabularx}{\linewidth}{@{}>{\RaggedRight\arraybackslash}p{0.06\linewidth}
>{\RaggedRight\arraybackslash}p{0.25\linewidth}
>{\RaggedRight\arraybackslash}p{0.40\linewidth}Y@{}}
\toprule
Version & Diagnostic question & Workflow action & Screening outcome \\
\midrule
V1 &
Can geometric displacement caused by skirt penetration alone explain plug
heave? &
Established a geometric baseline from solid-volume conservation
and suction-caisson design relations \cite{owa2019suction}. &
Produced the largest final-heave and process final-point errors among the 6
versions. \\

V2 &
How does suction-driven seepage change effective stress and void ratio? &
Added source-supported relations linking seepage, effective stress, and void ratio to update the soil state
\cite{tran2005seepage,xie2020seepage,been1985state}. &
Both errors decreased only slightly relative to V1. \\

V3 &
How should lateral effective stress and soil rebound evolve during confined
unloading? &
Updated lateral and mean effective stresses using relations for earth pressure at rest, stress history,
and granular-soil elasticity \cite{jaky1944k0,mayne1982k0ocr,suwal2013poisson}. &
Both errors decreased only modestly relative to V2. \\

V4 &
Does stress-path evolution produce additional void-ratio change in dense sand?
&
Introduced a stress-path-dependent void-ratio update based on state-parameter and state-dependent dilatancy relations \cite{been1985state,li2000dilatancy,jefferies1993norsand}. &
Produced the first substantial reduction in both screening errors relative to V3.
\\

V5 &
Can a more compact void-ratio update improve both final-state and process-level predictions? &
Tested an alternative void-ratio update under the same parameter-source rules. &
Reduced the process final-point error but increased the final-heave error.
\\

V6 &
Can the admitted mechanisms be assembled to keep both final-state and process errors low? &
Combined the audited seepage and void-ratio relations with
near-wall soil dilation and consistent final-state updating
\cite{bolton1986strength,ragni2020observations,houlsby2005design}. &
Retained a low final-heave error and achieved the lowest process
final-point error; V6 was selected. \\
\bottomrule
\end{tabularx}
\end{table}

The first cycle expanded the model beyond geometric displacement to account for
suction-induced changes in soil state. V1 used solid-volume conservation to
calculate the geometric heave caused by skirt penetration but did not account
for changes in soil void state under suction. The marked discrepancy from the measured plug heave indicated that geometric displacement alone
could not explain the observed plug heave response. Therefore, the workflow focused the next diagnostic question on how
suction-induced seepage changes the effective stress and void ratio of the soil
inside the caisson. To address this
question, the knowledge engine retrieved source-supported relations linking
suction, seepage, effective stress and void ratio \cite{tran2005seepage,xie2020seepage,been1985state}, and assembled them into V2.

V2 altered the predicted plug heave
only marginally. An audit of the update chain found that the
vertical effective stress responded to hydraulic loading, while the lateral
effective stress remained tied to a fixed stress ratio. Since the soil column
inside the caisson is laterally constrained, the workflow next examined the evolution of lateral effective stress during vertical unloading.
Relations describing earth pressure at rest,
stress history, and the elastic response of granular soils were incorporated into V3, which updated the lateral and mean effective stresses along a
confined-unloading path \cite{jaky1944k0,mayne1982k0ocr,suwal2013poisson}. This revision reduced the final-heave error only modestly.

Final-heave mean absolute percentage error (MAPE) remained above 70\% for
V1--V3 (Figure~\ref{fig:candidate-screening}), indicating that the represented volumetric response remained insufficient to reproduce the
measured plug heave. The audit
localized the major deficiency to the conversion of stress change into
void-ratio change. Then, the second diagnostic cycle examined how the distance from the
critical state bounds the available volumetric change in dense sand and how
movement along the stress path mobilizes that change. The knowledge engine
combined the resulting state-parameter and state-dependent dilatancy relations
to construct the stress-path-dependent void-ratio update in V4. This revision produced the first
substantial improvement in both screening metrics: final-heave and process
final-point MAPE fell to 12.3\% and 10.3\%, respectively
(Figure~\ref{fig:candidate-screening}).

The third diagnostic cycle tested whether final-state and process-level performance could be improved simultaneously. Under the same parameter-source rules, V5 tested a more compact void-ratio update. Its process final-point MAPE decreased to 6.6\%, but its
final-heave MAPE increased to 15.0\%, revealing a trade-off between the two
evaluation targets. This trade-off motivated V6, which combined the audited
seepage and void-ratio relations with near-wall soil dilation and consistent
final-state updating. V6 achieved a final-heave MAPE of 12.4\% and reduced the
process final-point MAPE to 4.2\%. Its final-heave performance remained comparable to that of V4, while its process final-point error was the lowest among the 6 formulations. This balance supported the selection of V6 for subsequent analysis.

After the physical knowledge and model inputs had been prepared, development of
V1--V6 required 171.5 min (2.9 h) of total agent execution time
(Table~\ref{tab:development-time}). Of this total, derivation required 42.5 min,
implementation required 82.5 min, and independent audit required 46.5 min, with implementation accounting for the largest share. Execution time for an individual formulation ranged from
19.3 min for V1 to 37.5 min for V4. 
% These timings cover the agent work from
% staged modeling questions to independently audited solver versions.
Across all 6 formulations, the recorded agent used 73.07 million tokens, corresponding to an estimated API-equivalent cost of US\$42.23 under the GPT-5.4 standard rates available in June 2026. These measurements cover the agent execution required to translate staged modeling questions into implemented and independently audited solver versions; they do not include the preceding preparation of physical knowledge and model inputs or the time required for human review.

\begin{table}[!t]
\centering
\caption{Agent execution time, token use, and estimated API-equivalent cost for
the V1--V6 model-development.}
\label{tab:development-time}
\footnotesize
\begin{tabular*}{\linewidth}{@{\extracolsep{\fill}}lcccccc@{}}
\toprule
Version &
\shortstack{Derivation\\(min)} &
\shortstack{Implementation\\(min)} &
\shortstack{Independent\\audit (min)} &
\shortstack{Total\\(min)} &
\shortstack{Tokens\\(million)} &
\shortstack{API equivalent \\cost (US\$)} \\
\midrule
V1 & 4.1  & 8.5  & 6.7  & 19.3 & 2.96  & 2.03 \\
V2 & 10.8 & 17.6 & 7.4  & 35.7 & 5.14  & 3.82 \\
V3 & 6.0  & 8.6  & 7.3  & 21.9 & 8.79  & 5.57 \\
V4 & 5.8  & 21.2 & 10.5 & 37.5 & 15.44 & 8.70 \\
V5 & 6.3  & 7.5  & 7.4  & 21.2 & 18.01 & 9.72 \\
V6 & 9.4  & 19.1 & 7.3  & 35.8 & 22.72 & 12.39 \\
Total & 42.5 & 82.5 & 46.5 & 171.5 & 73.07 & 42.23 \\
\bottomrule
\end{tabular*}

\medskip
\parbox{\linewidth}{\footnotesize\raggedright Execution time is the sum of the derivation, implementation, and independent-audit role durations recorded after the physical knowledge and model inputs had been prepared. Token totals comprise input and
output tokens across the corresponding recorded agent turns; cached input is counted once within input, and reasoning tokens are included within output. API-equivalent costs use the GPT-5.4 standard rates available in June 2026: US\$2.50, US\$0.25, and US\$15.00 per million uncached-input, cached-input, and output tokens, respectively. \par}
\end{table}

% --------------------------------------
\subsection*{Model selection remains stable across most study exclusions}
\label{retrospective-selection-stability}

\begin{figure}[h!]
\centering
\includegraphics[width=0.95\textwidth]{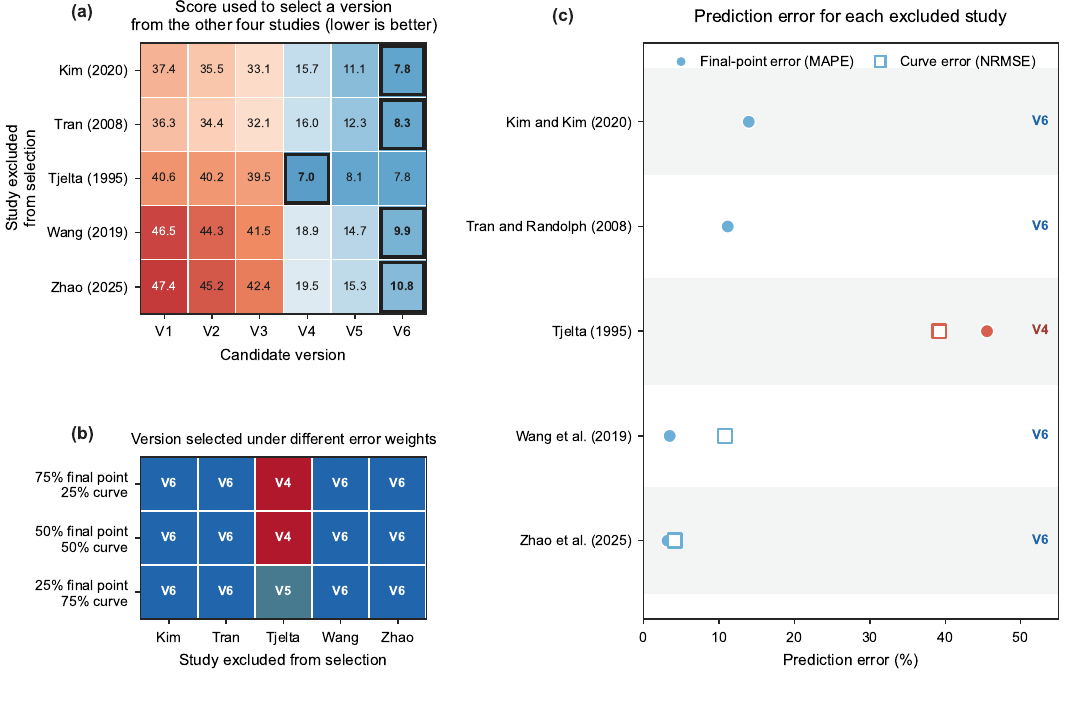}
\caption{Retrospective leave-one-study-out assessment of candidate-selection stability. The 6 candidate formulations were fixed before the analysis.
In each fold, one study was excluded from candidate scoring and selection, and
the selected formulation was subsequently evaluated against the held-out study.
(a) Scores of V1--V6 under the primary endpoint-to-curve weighting
of 75:25; boxes denote the selected lowest-scoring formulations.
(b) Formulations selected using endpoint-to-curve error weights
of 75:25, 50:50, and 25:75.
(c) Held-out endpoint MAPE and curve normalized root-mean-square
error (NRMSE) of the selected formulation.}
\label{fig:loso-selection-stability}
\end{figure}

After screening against the full literature pool selected V6, we
performed a five-fold retrospective leave-one-study-out analysis to test
whether this selection was dominated by an individual source study or by the weighting assigned to endpoint and process-curve errors. The 6 candidate formulations were fixed before this analysis. In each fold, one study was excluded, V1--V6 were rescored using
the remaining four studies, and the selected version was evaluated against the
held-out study. The analysis was repeated using endpoint-to-curve error weights
of 75:25, 50:50, and 25:75. Figure~\ref{fig:loso-selection-stability}(a) presents
the candidate scores under the primary 75:25 weighting,
Figure~\ref{fig:loso-selection-stability}(b) compares the selections across all
three weightings, and Figure~\ref{fig:loso-selection-stability}(c) reports the
held-out errors of the selected versions.

Under the primary weighting, V6 was reselected in four of the five folds
(Figure~\ref{fig:loso-selection-stability}(a)). The same four folds selected V6 under all three weighting schemes (Figure~\ref{fig:loso-selection-stability}(b)), and
their held-out endpoint MAPEs ranged from 3.2\% to
13.9\%
(Figure~\ref{fig:loso-selection-stability}(c)). Selection changed only when the
Tjelta study~\cite{tjelta1995geotechnical} was excluded. In this fold, the 75:25 and 50:50 weightings selected V4, whereas the 25:75 weighting, which placed greater emphasis on the process curves, selected V5 (Figure~\ref{fig:loso-selection-stability}(b)). Tjelta provides the only
full-scale field heave--depth profile in the evidence pool. Its exclusion left the
candidate scores primarily determined by centrifuge and laboratory model tests,
which favored V4 under endpoint-dominated evaluation and V5 under curve-dominated
evaluation. 

When evaluated against the held-out Tjelta profile, the selected V4 produced an
endpoint error of 45.5\% and a curve normalized root-mean-square error (NRMSE) of
39.2\%. For comparison, V6 produced corresponding errors of 8.2\% and 20.4\%, respectively. 
Overall, candidate selection remained stable across four of the five study exclusions and all tested weighting schemes within those four folds, but changed when the only full-scale study was removed. The relatively lower held-out errors of V6 for the Tjelta profile indicate that it provided a more consistent balance across centrifuge tests, laboratory model tests, and the available full-scale field response, supporting its retention as the selected formulation.

%=============================================================================
\subsection*{The selected model integrates three plug-heave mechanisms}
\label{soil-plug-model-structure}

\begin{figure}[!t]
\centering
\includegraphics[width=0.75\textwidth]{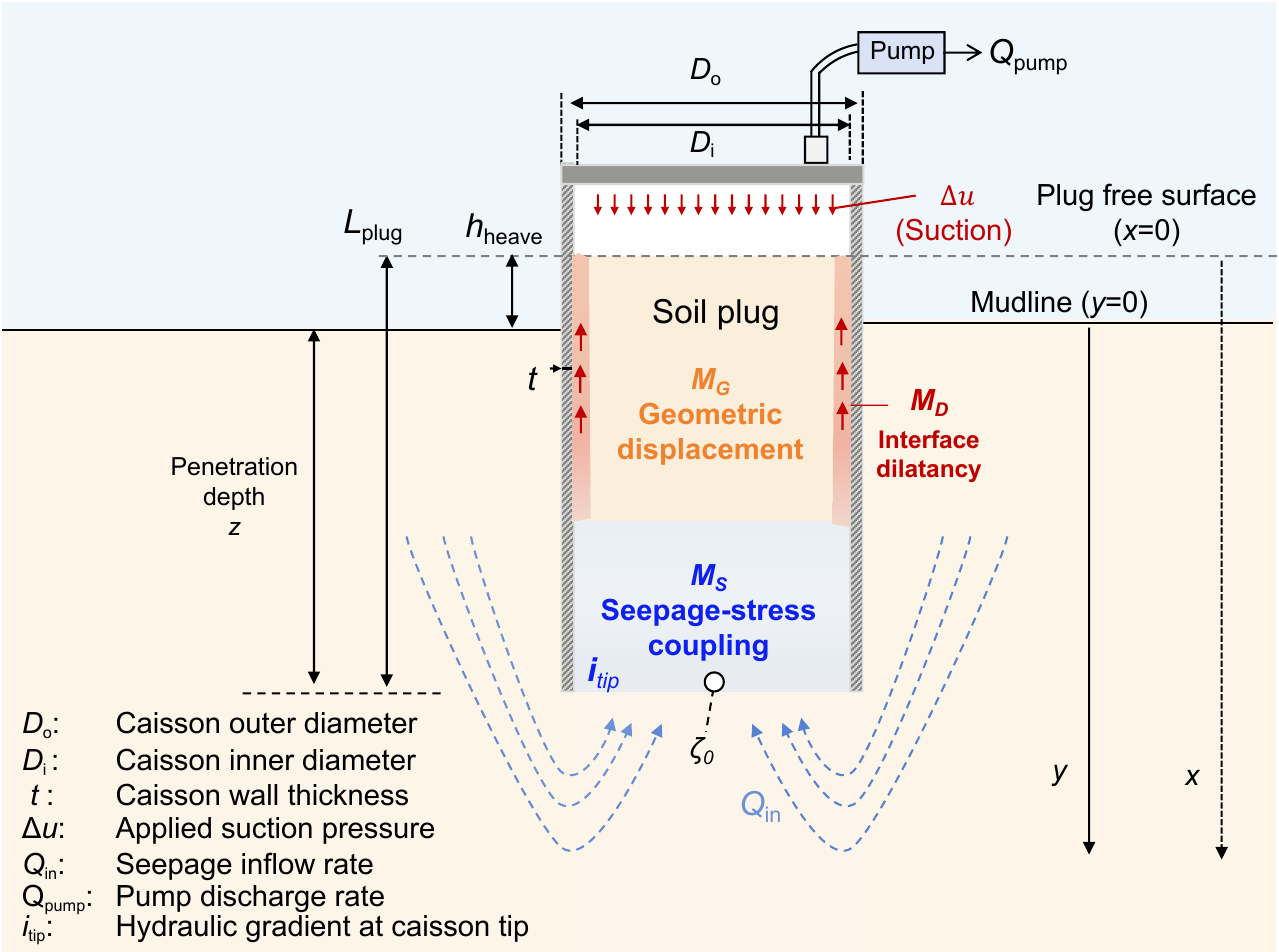}
\caption{Physical structure and coordinate definitions of the selected soil-plug formulation. The fixed
coordinate $y$ is measured downward from the initial seabed, and the local
coordinate $x$ is measured downward from the current plug surface. At
penetration depth $z$, the plug length is
$L_{\mathrm{plug}}=z+h_{\mathrm{heave}}$. The initial material coordinate
$\zeta_0$ tracks the soil entering the internal control volume. $M_G$, $M_S$,
and $M_D$ denote skirt-penetration-induced geometric displacement,
seepage-driven soil void-ratio evolution, and near-wall soil dilation,
respectively. Dashed blue curves schematically indicate the seepage paths.}
\label{fig:soil-plug-model-structure}
\end{figure}

The selected V6 formulation predicts plug heave by combining skirt-penetration-induced geometric
displacement ($M_G$), seepage-driven soil void-ratio evolution ($M_S$), and
near-wall soil dilation ($M_D$) (Figure~\ref{fig:soil-plug-model-structure}).

At penetration depth $z$, these three contributions are assembled through the solid-volume-conservation relation:

\begin{equation}
h_{\mathrm{heave}}(z)
=
\int_0^z
\left[
\alpha_A
\frac{1+e(\zeta_0,z)}{1+e_0}
-1
\right]
\,\mathrm{d}\zeta_0,
\label{eq:results-plug-heave}
\end{equation}
where $\zeta_0$ labels a soil layer by its initial depth below the seabed, and
$z$ denotes the current penetration depth. The quantity $e(\zeta_0,z)$ is
the void ratio of that soil layer after the $M_S$ update and subsequent $M_D$
correction at penetration depth $z$. The initial void ratio is $e_0$, and
$\alpha_A=(D_o/D_i)^2$ is the caisson area ratio, where $D_o$ and $D_i$ are the
outer and inner caisson diameters, respectively.

When no void-ratio change occurs, $e(\zeta_0,z)=e_0$, and
Equation~\eqref{eq:results-plug-heave} reduces to the geometric limit:

\begin{equation}
h_{\mathrm{heave}}(z)=(\alpha_A-1)z,
\label{eq:results-geometric-heave}
\end{equation}

Equation~\eqref{eq:results-geometric-heave} recovers the geometric contribution
$M_G$. In the full formulation, $M_S$ and $M_D$ modify the predicted plug heave through their
updates to $e(\zeta_0,z)$, while Equation~\eqref{eq:results-plug-heave} assembles these contributions into the observable response in V6. The coupled seepage,
effective-stress, and void-ratio relations used to obtain $e(\zeta_0,z)$ are
provided in Supplementary Note 1. Supplementary Fig. 1 presents the sequence through which these
component calculations are evaluated and assembled at each penetration step.

%=============================================================================
\subsection*{Multi-agent audits recover known and previously unflagged issues}
\label{workflow-level-verification}

The blinded replay tested whether the retained development records contained
enough information for an isolated audit role to recover known modeling issues. With the original discussions and expected answers withheld, the audit role recovered and
localized all 9 issues, reported no issues outside the target set, and
completed the scored replay without human assistance
(Table~\ref{tab:workflow-verification}). These results show that
the retained records were sufficient to identify where each known issue arose
and which physical rule, computational operation, or code segment required re-examination.

The derivation-to-code probes then examined whether a separate audit role could detect
implementation problems missed by the predefined checks. Across the 5 probes, the implementation role passed all 36 fixed checks. 
The audit role then reported five additional findings across 4 of the 5
probes (Table~\ref{tab:workflow-verification}). These findings concerned an omitted rule
for refreshing the final output, branch execution,
the scope of a stated rule, an incomplete audit trace, and a numerical safeguard
applied to the wrong term in an equation. This indicates that the role-separated audit extended the evaluation beyond the coverage of the predefined checks, recovering an omitted requirement and identifying implementation issues that those checks did not address.

After these findings had been addressed, the final code used for benchmark
evaluation passed all 37 code tests, comprising 33 module-level tests and 4 regression tests (Table~\ref{tab:workflow-verification}). It also passed all 5 independently calculated reference checks. Within the scope of these tests, the results confirmed that the final
implementation followed the equations and calculation sequence described in
the manuscript. The blinded-replay and derivation-to-code probe records are provided in Supplementary Note 2, and the final-code tests and regression baselines are reported in Supplementary Note 3.

\begin{table}[!t]
\centering
\small
\caption{Results of workflow audit and final code verification.}
\label{tab:workflow-verification}
\begin{tabularx}{\textwidth}{
@{}>{\raggedright\arraybackslash}p{0.18\textwidth}
 >{\raggedright\arraybackslash}p{0.4\textwidth}
 >{\raggedright\arraybackslash}X
 @{}}
\toprule
Audit task & Observed result & Supported conclusion \\
\midrule
Blinded replay of known issues &
All \textbf{9} known issues were recovered and localized; no issues outside the target set, and no human assistance was provided during the scored replay. &
The retained development records enabled an isolated audit role to recover and localize known issues under blinded conditions. \\
Derivation-to-code probe audit &
All \textbf{36} predefined checks passed; the separate audit reported \textbf{5} additional findings across \textbf{4} of the \textbf{5} probes, with no finding in one
probe. &
The role-separated audit identified both an omitted predefined rule and implementation
issues not covered by the predefined checks. \\
Final code verification &
All \textbf{37} code tests passed, comprising \textbf{33} module-level tests
and \textbf{4} regression tests; all \textbf{5} independently calculated reference checks
also passed. &
Within the checked scope, the final implementation followed the equations and
calculation sequence described in the manuscript. \\
\bottomrule
\end{tabularx}
\end{table}

%----------------··············

\subsection*{Staged benchmarks support final-state and process-level predictions}
\label{staged-benchmark-predictions}

\begin{figure}[p]
\centering
\includegraphics[ width=\linewidth,
  height=0.7\textheight,
  keepaspectratio]{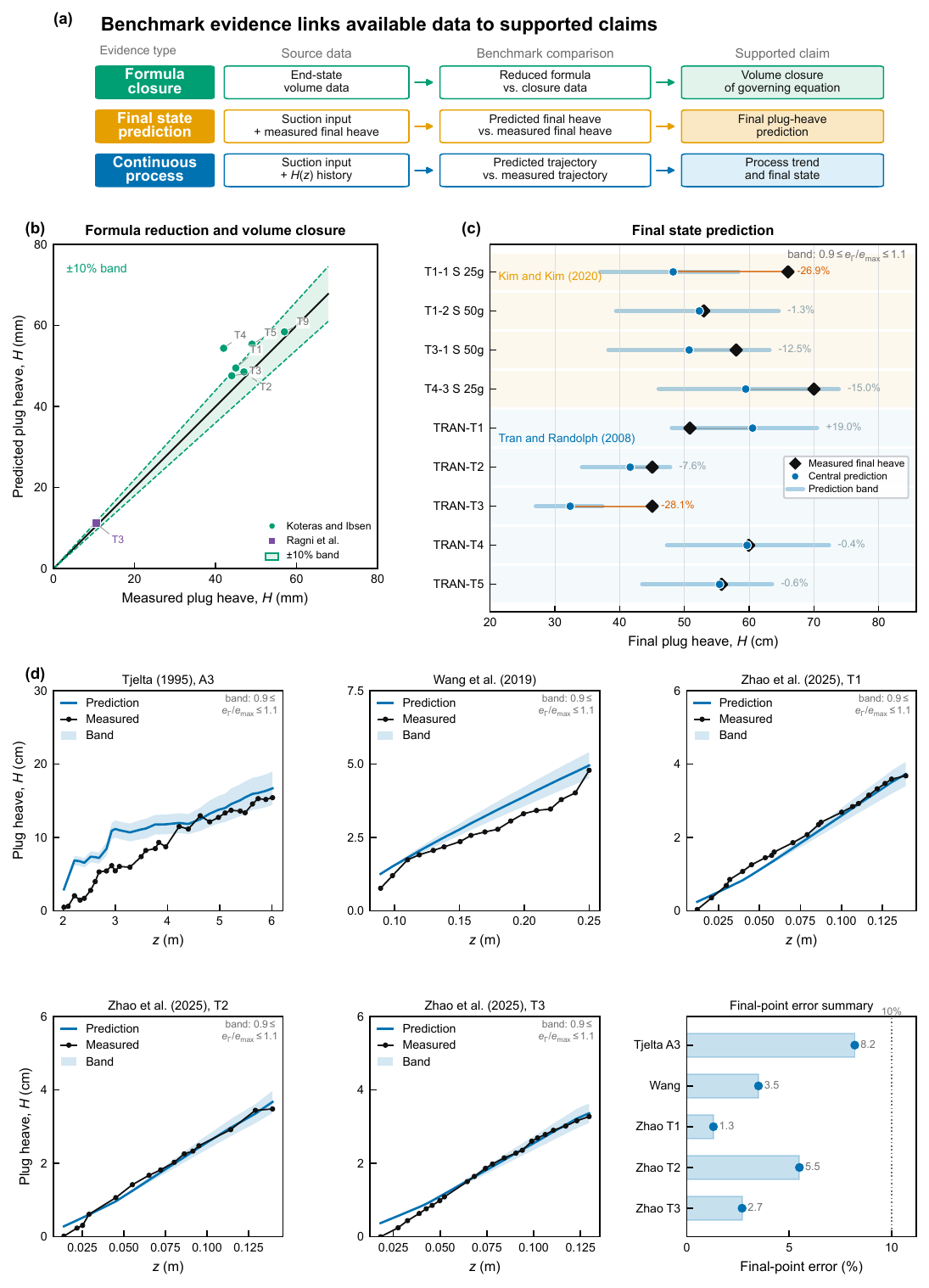}
\caption{Staged benchmark evaluation of V6 plug-heave predictions. (a) Evidence map linking the available source data, benchmark problem, and supported claim at each evaluation level. (b) Formula reduction and volume closure for 7 cases using source-reported changes in void ratio or volumetric strain. The diagonal line represents equality between predicted and measured final heave, and the shaded region denotes the predefined ±10\% comparison band. (c) Final state prediction for 9 benchmark cases. Black diamonds are measured final heave, blue points represent central predictions, and horizontal bands show the sensitivity range obtained by varying $e_{\Gamma}/e_{max}$ from 0.9 to 1.1. Percentages report the signed errors of the central predictions. (d) Predicted and measured plug-heave–depth histories for 5 cases, together with the corresponding sensitivity bands and final-point errors. The bands in panels (c) and (d) represent prescribed parameter sensitivity rather than probabilistic confidence intervals.}
\label{fig:benchmark-validation}
\end{figure}

We evaluated V6 using three levels of plug-heave
evaluation: formula reduction and volume closure, final state prediction, and
continuous process prediction
(Figure~\ref{fig:benchmark-validation}(a)).

At the first level, final plug heave was calculated using reported changes in void
ratio or volumetric strain. 6 of the 7 V6 calculations fell within or close to the predefined $\pm 10\%$
comparison band (Figure~\ref{fig:benchmark-validation}(b)). 6 cases used
void-ratio changes reported by Koteras and
Ibsen~\cite{koteras2019medium}, while the Ragni et
al.~\cite{ragni2020observations} T3 case used an image-derived volumetric
strain. Only T4 showed a clear deviation, with an error of $+29.5\%$. This difference may reflect experimental scatter or uncertainty in the
reported measurement: T4 and T5 had the same void-ratio change and similar
penetration depths, but their measured final plug heaves differed by 7\,mm,
equivalent to 16.7\% of the T4 measurement. The Ragni T3 case yielded an error
of $+6.6\%$, showing that the volume-conservation relation also reproduced the measured final heave when
volumetric strain was provided directly.

At the second level, V6 predicted final plug heave for each case from the
initial soil state and suction-pressure history. The prediction sensitivity bands
encompassed 7 of the 9 benchmark endpoints, and the central predictions
yielded a MAPE of $12.4\%$
(Figure~\ref{fig:benchmark-validation}(c)). The endpoints for Kim and
Kim~\cite{kim2020soil} T1-1 S 25g and Tran and
Randolph~\cite{tran2008variation} TRAN-T3 lay outside these bands. The upper
prediction bounds from the predefined sensitivity analysis reduced the
respective shortfalls from $26.9\%$ to $11.7\%$ and from $28.1\%$ to $17.1\%$, respectively. The reference endpoint for TRAN-T3 was inferred from the final digitized penetration depth under a plug-filled endpoint assumption, which introduced additional uncertainty into the endpoint comparison. Overall, V6 reproduced final plug heave for most benchmark cases, while the sensitivity analysis reduced the underprediction in the two outside-band cases.

The third level extended the prediction from penetration endpoints to complete
plug-heave--depth histories. Across 5 cases, V6 captured the overall increase in plug
heave with penetration, and all 5 measured endpoints lay within the
predefined prediction sensitivity bands
(Figure~\ref{fig:benchmark-validation}(d)). Final-point errors ranged from
$1.3\%$ to $8.2\%$, giving a MAPE of $4.2\%$. V6 agreed most closely with the
three curves reported by Zhao et al.~\cite{zhao2025formation}, with curve
normalized root-mean-square errors (NRMSEs) of $3.9\%$--$4.3\%$. The Wang et
al.~\cite{wang2019installation} and Tjelta~\cite{tjelta1995geotechnical} cases
yielded curve NRMSEs of $10.8\%$ and $20.4\%$, respectively. 

The largest process-level discrepancy occurred during early penetration in the Tjelta case, for which V6
predicted a faster increase in plug heave than was measured. This early
overprediction is consistent with the treatment of geometric displacement in V6: soil
displaced by skirt penetration contributes immediately to the calculated plug
heave, while under the relatively low suction present during early penetration, some displaced soil
may first redistribute before producing measurable heave. As penetration
continued, the predicted and measured responses converged and reached comparable
final magnitudes.

Together, the three benchmark levels support distinct aspects of V6 rather than a single undifferentiated validation claim. The first evaluates volume closure when the end-state soil response is supplied, the second evaluates final-heave prediction from the initial state and suction history, and the third evaluates the evolution and final magnitude of plug heave.
Detailed case results are
reported in Supplementary Table 13 for formula reduction and volume closure, Supplementary Table 14 for final state prediction, and Supplementary Tables 15 and 16 for continuous process prediction.

%=============================================================================
\phantomsection
\section*{Discussion}\label{discussion}

We used a multi-agent workflow operating within human-defined physical
boundaries to connect source-supported knowledge, theory construction, numerical implementation, audit, and benchmark evaluation. Iteration through 6 candidate formulations
produced V6, which integrates skirt-penetration-induced geometric displacement,
seepage-driven soil void-ratio evolution, and near-wall soil dilation through
solid-volume conservation. Its central predictions yielded a MAPE of 12.4\% across 9 final-state benchmark cases and a
final-point MAPE of 4.2\% across 5 measured
plug-heave--penetration-depth histories. With these, this study produced two linked outcomes: a physically interpretable model of soil-plug evolution and a traceable route from its knowledge
sources to its numerical implementation and evaluation.

This study extends agent-assisted scientific computing from isolated retrieval, coding, or data-analysis tasks to a longer
engineering-modeling chain. Existing multi-agent systems have generated
computational strategies within predefined scientific machine-learning
problems or linked literature-grounded hypotheses to experimental data analysis
\cite{jiang2026agenticsciml,ghareeb2026robin}. In the present study, no complete soil-plug model was available at
the outset. In this case, the workflow had to identify relevant physical
mechanisms, derive their governing relations, discretize the equations, and
implement the resulting solver. Human experts retained authority over admissible knowledge sources, modeling assumptions, and model revisions. When benchmark evaluation revealed a
substantial discrepancy between predicted and measured plug heave, or a
trade-off between final-heave and process errors, the agents formulated the
next physical question and retrieved relevant source-supported relations. Each admitted relation was then translated into equations, numerical operations, and code implementation. The methodological contribution therefore lies not in removing human judgment, but in organizing human and agent roles so that the complete modeling process remains inspectable.

The evolution from V1 to V6 illustrates how predictive discrepancies can be converted into physically meaningful diagnostic questions. V1--V3 continued to
underpredict plug heave after seepage and confined unloading had
been introduced, indicating that their stress updates generated insufficient
void-ratio change. The workflow consequently examined how movement along the stress path should be
converted into volumetric response. Introducing a state-dependent
volume-change relation in V4 substantially reduced both final-heave and process
final-point errors. V5 further reduced the process final-point error but
increased the final-heave error, revealing a trade-off between the two evaluation targets. V6 addressed this trade-off by integrating the admitted seepage, soil-state, and near-wall dilation mechanisms while maintaining consistent final-state updating. In this way,
each model revision was linked to a defined physical question, supporting evidence, a mathematical expression, and an implementation path.

The resulting formulation also provides a physical interpretation of plug-heave development.
Skirt penetration determines the volume of soil entering the caisson, while
the suction-induced stress path changes the soil state and near-wall dilation
further modifies the void-ratio field. Solid-volume conservation then
converts this evolving field into observable plug heave. Therefore, the plug heave can
be interpreted as a process response initiated by geometric
displacement and modulated by suction-driven soil-state evolution and interface behavior.

Complementary benchmark evidence supported the selection of V6. Its prediction sensitivity
bands encompassed 7 of the 9 final-state benchmark cases and the
endpoints of all 5 measured plug-heave--penetration-depth histories.
The retrospective leave-one-study-out analysis reselected V6 in 4 of the 5
folds and under all 3 scoring schemes in those folds. Selection changed only when the full-scale Tjelta
study~\cite{tjelta1995geotechnical} was excluded. The formulation selected without this field record produced substantially larger errors than V6 when evaluated against the held-out Tjelta profile. This comparison highlights the
importance of full-scale process data in constraining model selection and
supports the balance achieved by V6 across final-state, process-level, and
experimental-scale evidence.

From an engineering perspective, V6 extends the conventional geometric heave
estimate to a process-level prediction controlled jointly by the suction-pressure
history and evolving soil state. It retains a transparent geometric baseline while
describing how plug heave develops throughout penetration. The resulting
formulation provides a physically interpretable reduced-order model for
assessing both final plug heave and its development during suction-caisson installation.

Independent role-separated auditing addresses a different question from benchmark evaluation: whether the declared physical relations were translated consistently into equations and executable code. The blinded replay recovered and localized all 9 known problems
without access to the expected answers, showing that the retained records preserved the information required for subsequent examination. The derivation-to-code audit then identified 5 additional findings across 4 probes after all 36 predefined implementation checks had passed. This indicates that the separate audit role extended the coverage of the predefined checks by identifying an omitted requirement and cross-layer inconsistencies in the transition from derivation to implementation. Benchmark evaluation examined what the solver predicted, while the audit examined how the solver was constructed.

The evidence boundary of the current study is determined primarily by the available process data and the scope of the demonstration. Only 5 comparable plug-heave--penetration-depth histories are available,
including one full-scale field record, and few studies jointly report
suction-pressure history, continuous plug heave, and independently
measured soil state. V6 currently applies to vertical penetration in saturated
homogeneous sand and assumes pure soil inflow, quasi-steady Darcy seepage, and
cross-sectionally averaged near-wall soil dilation. More complete field
observations and explicit treatment of early soil redistribution,
stratification, and three-dimensional interface behavior will further test and
extend both the physical model and the workflow.

In summary, this study treats engineering-model development as an inspectable scientific process rather than an unreported precursor to the final solver.
The selected formulation (V6) demonstrates that this workflow can produce an engineering model with
explicit physical meaning, while benchmark evaluation and independent audit
examine its predictive outcome and formation process, respectively. The central
implication is that AI agents can participate in constructing engineering theory
within explicit human-defined physical boundaries, while the model's knowledge sources,
derivation path, numerical implementation, and predictive performance remain
open to scientific scrutiny.

%------------------------------------------

\phantomsection

%%----------——————————————
\section*{Methods}

\subsection*{Human-guided physics-constrained multi-agent workflow}

\begin{figure}[H]
\centering
\includegraphics[width=\textwidth]{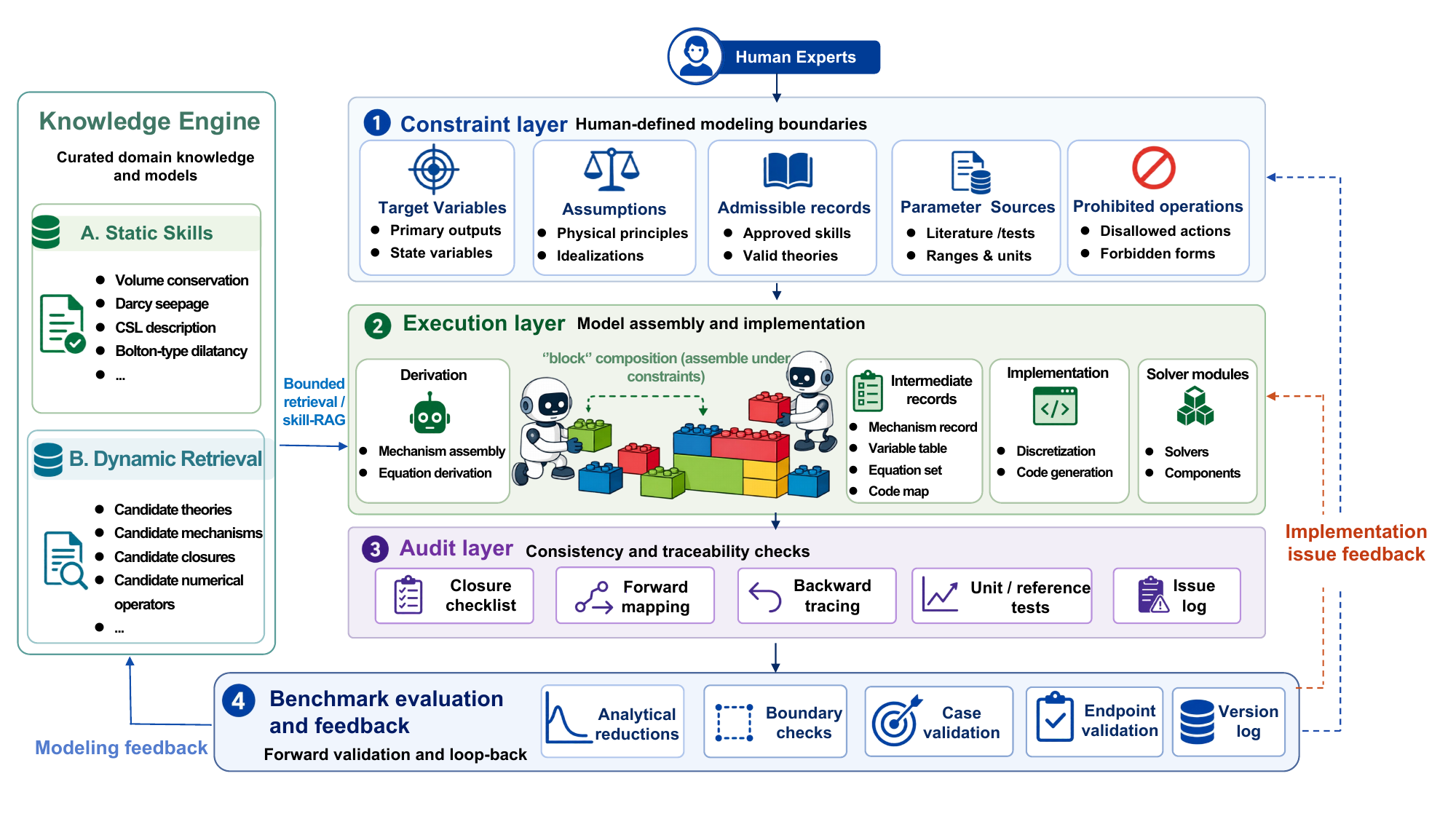}
\caption{Overview of the human-guided physics-constrained multi-agent workflow. Blue and orange dashed arrows indicate feedback on physical assumptions and implementation issues, respectively.}
\label{fig:workflow-architecture}
\end{figure}

The human-guided physics-constrained multi-agent workflow organizes model development into five linked stages: knowledge admission, skill composition, solver construction, independent audit, and benchmark evaluation (Fig.~\ref{fig:workflow-architecture}). These stages are implemented through the knowledge engine, constraint and skill-composition modules, execution layer, audit layer, and benchmark-validation interface.

The workflow is designed to identify and avoid errors that can enter the modeling chain before the solver is validated against benchmark problems. These errors mainly arise from three aspects:
\begin{itemize}
    \item What knowledge is admitted, e.g., an assumption used outside its
    admissible range.
    \item How agents use the knowledge, e.g., a symbol that changes meaning
    during derivation, or a temporary iteration value recorded as a final input.
    \item How solver outputs are verified, e.g., a failure to trace a reported
    prediction back to the physical relations and calculation steps that
    produced it.
\end{itemize}

Figure~\ref{fig:workflow-architecture} maps these critical aspects onto the
workflow structure: the knowledge engine supplies traceable physical records;
the constraint layer decides which physical laws and operations are admissible;
the execution layer turns the admitted components into equations and code under
fixed notation and state rules; and the audit layer checks consistency across
the physical rules, equations, and code and returns unresolved inconsistencies
for human review.
Throughout this sequence, human experts retain authority over record admission
and model revision, while agents perform bounded retrieval, derivation, code
implementation, and auditing tasks.

The first human decision occurs in the knowledge engine, the entry point for
physical information. It contains two categories of agentic skills that serve
as building blocks for constructing the theoretical framework.
\begin{itemize}
    \item Static skills: human-approved physical laws, constitutive
    relationships, or numerical operators. An agent can add a knowledge block
    to the model only when its assumptions, variables, and boundary conditions
    are compatible with those of its neighboring blocks. This bounded
    search-and-use strategy forms the skill-RAG operation of the workflow. The
    language model is guided by records retrieved during use
    \cite{lewis2020retrieval}, and the retrieval target is a task-relevant skill
    rather than unrestricted text \cite{wei2026skill}.
    \item Dynamic retrieval: if the initially defined static skills do not
    contain everything needed for model assembly, an agent may use the LLM's
    pretrained knowledge, web search, or other tools to locate candidate
    theories, published mechanisms, closure assumptions, or numerical
    operators \cite{yao2023react}. Before use, the LLM-identified item needs to
    be reframed as a compatible skill block. This process requires its source
    to be identified, its physical relation to be defined, and its allowable
    use to be stated, including variables, units, assumptions, and applicable
    range. A human reviewer then decides whether the item is admitted as a new
    skill or kept only as supporting evidence.
\end{itemize}

After records have been admitted, the execution layer controls how agents use
them to build the solver. Agents select compatible records, assemble
mechanisms, derive equations, and implement code modules under the approved
assumptions and parameter rules. The execution layer also fixes numerical
conventions before solver outputs are accepted, including coordinates, signs,
gradient directions, and the volumetric-strain convention. It separates trial
values from the accepted state, defined here as the checked version of
variables, inputs, and update history submitted for reporting. This rule
ensures that version records, committed histories, and reported outputs refer
to the same calculation iteration.

The audit layer then evaluates whether the continuous equations, discrete
update rules, and code modules remain consistent. If a mismatch is found, the
affected item returns to the execution layer for correction and is then checked
again. Agents can identify and localize these inconsistencies, but they cannot
admit new records or authorize model changes until human review. In this
implementation, the derivation, coding, and audit roles were run in OpenAI
Codex with GPT-5.4 as the underlying model. Codex performed bounded tasks,
while humans decided whether new records could be admitted and whether the
model should be revised.

The retained audit materials and results are documented in Supplementary Note 2.

The same loop applies after benchmark validation. After the audit layer establishes internal consistency, the model parameters remain fixed while differences between predicted and reported plug heave are evaluated against two predeclared bounds: the comparison band, which defines acceptable differences from the literature data, and the prediction sensitivity band, which represents prediction variation caused by documented input uncertainty. An outside-bound
prediction attributed to a missing mechanism or unsuitable assumption returns
the model to the knowledge engine, constraint layer, and human review. A difference attributable to documented source uncertainty, benchmark definition, or declared model scope is retained as a bounded modeling limitation.

%%----------——————————————
\subsection*{Soil-plug model-development task}

To apply the workflow, we defined soil-plug evolution during vertical
suction-caisson penetration in saturated, homogeneous sand as the
model-development task (Fig.~\ref{fig:fig1}(b)). The modeled system was the soil
entering the caisson and forming an internal plug during penetration, and the
target response was its rise relative to the surrounding seabed. Task inputs
comprised caisson geometry, initial soil state and material properties,
penetration depth, and the available suction-pressure history. From these
inputs, the agents were tasked with constructing an executable model to
calculate plug heave as penetration progressed and at the final penetration
depth.

At task initiation, the admitted static skills comprised solid-volume
conservation, Darcy-type seepage, critical-state-line (CSL) descriptions, and
Bolton-type dilatancy. These skills formed the initial knowledge base for model
development. When additional physical relations or closure assumptions were required during model development, they followed the knowledge-admission procedure described in the preceding subsection.

Development time was recorded for the derivation, implementation, and
independent audit roles after the physical knowledge and model inputs had been
prepared. For each version, role duration was the elapsed time required to
complete the corresponding output, and total agent execution time was the sum
of the three role durations.

%%----------——————————————
\subsection*{Workflow audit and implementation verification}

To evaluate whether the workflow could detect and localize problems along the model-development chain, we used two audit tasks followed by a final code check:

\begin{itemize}
\item Blinded audit replay: tested whether an audit role could recover known issues using only the retained development materials.

\item Derivation-to-code probes: tested whether a separate audit role could identify problems not covered by the fixed checks after code generation.

\item Final code verification: checked whether the corrected code used for benchmark evaluation followed the equations and calculation sequence described in the manuscript.

\end{itemize}

In the blinded audit replay, the audit role received development materials but
was not given the target issue type, location, or answer key. The task tested
whether the role could recover and localize the issue from those materials
alone. We first converted 9 consistency issues identified during workflow
execution into audit packets that contained no answers, while retaining the
corresponding answer keys separately for scoring. Each packet contained only
the relevant physical relation, computational rule, and pre-correction code.
An isolated audit role then reviewed one packet using a general audit
instruction, without access to issue labels, answer keys, prior discussions,
or any material outside the packet. Finally, we compared each audit report
with the corresponding answer key and scored whether the role detected and
localized the issue, explained its consequences, and proposed a valid
correction or follow-up check. False positives and human assistance were
recorded separately.

The derivation-to-code probes extended the assessment to code generation. We
constructed 5 controlled probes. For each probe, an implementation role generated code from the supplied equations and state-update rules, followed by the corresponding fixed checks. Across the 5 probes, 36 predeclared checks were completed in total. A separate audit role then received the derivation materials, generated code, and check records. It assessed equation-to-code consistency, focusing on boundary and branch handling, the timing of state and output updates, and the completeness of the check records. Full procedures for preparing the audit materials, limiting the information provided to each role, and scoring the reports are described in Sections S2.1--S2.2 of Supplementary Note 2.

After the audit findings had been addressed, the code used for benchmark evaluation was checked for consistency with the governing equations and calculation sequence. The check covered key calculations, repeated runs with fixed inputs, and comparisons between key outputs and independently calculated reference values. The complete test list and passing criteria are provided in Supplementary Note 3.

%--------------------------------------
\subsection*{Benchmark evaluation and model selection}

\subsubsection*{Benchmark datasets}
For the soil-plug case study, 3 benchmark problems were formulated according to the type of suction-response information available in the literature:	reported end-state volumes, final plug heave, or complete heave--depth histories.
	\begin{itemize}
	    \item Formula reduction and volume closure were evaluated using 7 cases
	    from 2 studies~\cite{koteras2019medium,ragni2020observations}. Void
	    ratios or volumetric strains before and after suction installation were
	    provided to calculate the final plug heave directly from the end-state
	    volumes and to verify the overall volume closure.
	    \item Final state prediction was evaluated using 9 cases from 2
	    studies~\cite{kim2020soil,tran2008variation}. The pre-suction soil void
	    ratio, the suction-pressure history, and the test conditions were
	    provided to predict the final plug heave and the associated prediction
	    sensitivity band.
	    \item Continuous-process prediction was evaluated using 5 cases with
	    measured heave--depth histories from 3
	    studies~\cite{tjelta1995geotechnical,wang2019installation,zhao2025formation}.
	    The initial soil state and the suction-pressure history were provided to
	    predict the development of plug heave during penetration.
	\end{itemize}

\subsubsection*{Input and parameter assignment}
Benchmark inputs were assigned through a fixed source hierarchy.
Validation targets were obtained from reported measurements or
reconstructed from published figures. Values explicitly reported in the source studies
were used first: measured quantities (direct) and test-design parameters
(design parameter). Inputs not reported explicitly were derived where
possible from reported information or physical relationships (derived). The remaining inputs were assigned from data for the same material (analog) or from predefined engineering assumptions (assumption). Each adopted value and its source class were recorded in a parameter-source ledger (Supplementary Data 2). The label unavailable identifies entries for which no sourced value was adopted; the label mixed identifies rows combining values from multiple source classes. All inputs were fixed before any model calculation, so that no input could be adjusted in response to prediction outcomes. Each candidate formulation was then evaluated by comparing its predictions with the validation targets. Supplementary Note 4 documents the inputs, source labels, and derivations for each benchmark case, and the underlying numerical records are provided as Supplementary Data 1.

Most benchmark inputs were taken directly from the source studies or derived through the hierarchy above. Three model parameters, however, were assigned by common rules in the final-state and continuous-process predictions. None of the five prediction studies provided case-specific values of the critical-state-line intercept ratio \(e_{\Gamma}/e_{\max}\) or Poisson's ratio \(\nu\); the central value of \(e_{\Gamma}/e_{\max}\) was therefore set to 1.0, the midpoint of the sensitivity range adopted
below, and \(\nu\) was set to 0.30~\cite{suwal2013poisson,thota2021poisson}. Kim and
Kim~\cite{kim2020soil} reported case-specific \(k_i/k_{\mathrm{out}}\)
values ranging from approximately 1.26 to 1.46, and a representative
value of 1.3 was adopted for their cases; the remaining cases used
\(k_i/k_{\mathrm{out}}=3.0\), following Houlsby and
Byrne~\cite{houlsby2005design}. Formula reduction and volume closure
required no such assignments, because their calculations used the
source-reported end-state volumes directly.

Sensitivity to \(e_{\Gamma}/e_{\max}\) was evaluated by varying it by \(\pm 10\%\) about the central value, giving a range of 0.9--1.1; the adopted range was based on reported critical-state-line and state-parameter data for
sand~\cite{woo2019numerical,zhu2020parameter,liu2024evaluation,kurniadi2025drained}.
The resulting spread defines the prediction sensitivity band for the final-state and continuous-process predictions. This band isolates the effect of \(e_{\Gamma}/e_{\max}\): no case-specific values of this ratio were available for the prediction studies, and it sets the low-stress reference used in the void-ratio update, whereas Poisson's ratio and the permeability ratio were held at the values assigned above.

\subsubsection*{Evaluation metrics}

The metrics were chosen to match the validation data available for each
benchmark. For formula reduction and volume closure, final plug heave was
calculated from the reported void-ratio or volumetric-strain information and
compared with the reported final plug heave. Agreement was assessed against the
predefined comparison band of \(\pm 10\%\).

For final state prediction, we expressed the signed percentage error for
benchmark case \(c\) as

\begin{equation}
E_c =
\frac{h_{\mathrm{pred},c}-h_{\mathrm{meas},c}}
{h_{\mathrm{meas},c}}\times 100\%,
\label{eq:benchmark-case-error}
\end{equation}

where \(h_{\mathrm{pred},c}\) is the predicted final plug heave and
\(h_{\mathrm{meas},c}\) is the corresponding final plug-heave value used for
validation. The value of \(h_{\mathrm{meas},c}\) was either reported directly or reconstructed from the final digitized penetration depth for that case, as described above. A positive \(E_c\) indicates overprediction, whereas a negative
value indicates underprediction. We then summarized the absolute case-level
errors as the mean absolute percentage error:

\begin{equation}
\mathrm{MAPE}
=
\frac{1}{N_c}
\sum_{c=1}^{N_c}\left|E_c\right|,
\label{eq:benchmark-final-mape}
\end{equation}

where \(N_c\) is the number of cases included in the final state prediction
average.

For continuous process prediction, the predicted plug heave was evaluated at
the measured penetration depths. The error at the final measured depth was
computed using Equation~\eqref{eq:benchmark-case-error}. Averaging its absolute
value across the process cases gives the final-point MAPE. Because percentage
error is undefined when the measured heave is zero, curve MAPE was calculated
only at points with nonzero measured plug heave:

\begin{equation}
\mathrm{MAPE}_{\mathrm{curve}}
=
\frac{1}{N_{+}}
\sum_{p\in\mathcal{P}_{+}}
\left|
\frac{h_{\mathrm{pred},p}-h_{\mathrm{meas},p}}
{h_{\mathrm{meas},p}}
\right|
\times 100\%,
\label{eq:benchmark-curve-mape}
\end{equation}

where \(p\) indexes the process points, and \(h_{\mathrm{pred},p}\) and
\(h_{\mathrm{meas},p}\) are the predicted and measured plug heave at point
\(p\), respectively. The set \(\mathcal{P}_{+}\) contains the points for which
both values are finite and \(h_{\mathrm{meas},p}\neq 0\), and
\(N_{+}=|\mathcal{P}_{+}|\).

To complement this relative measure, we calculated the root-mean-square error
in physical units over all process points with finite predicted and measured
plug heave, including points with zero measured heave:

\begin{equation}
\mathrm{RMSE}_{\mathrm{curve}}
=
\left[
\frac{1}{N_p}
\sum_{p\in\mathcal{P}_{f}}
\left(
h_{\mathrm{pred},p}-h_{\mathrm{meas},p}
\right)^2
\right]^{1/2},
\label{eq:benchmark-curve-rmse}
\end{equation}

where \(\mathcal{P}_{f}\) is the set of process points for which both the
predicted and measured plug heave are finite, and \(N_p=|\mathcal{P}_{f}|\).
Plug heave is expressed in centimeters.

\subsubsection*{Candidate screening}
Each candidate formulation generated by the workflow was evaluated using the
same benchmark cases, input assignments, and parameter rules. Because the
candidate formulations shared the same volume-closure relation, comparison
during model development focused on final state prediction and continuous
process prediction. Final-heave MAPE characterized agreement with the
endpoint data, whereas process final-point MAPE characterized agreement at the
end of the measured penetration histories. Both metrics were retained in the
development records and used to determine whether the formulation entered
another cycle of audit and revision.

\subsubsection*{LOSO selection stability}

The dataset used for candidate selection contained 5 source studies. Each study was held out once, giving 5 retrospective leave-one-study-out (LOSO) folds. In each fold, the remaining 4 studies were used to rescore the candidate models and identify the best-performing model, which was then evaluated against the held-out study. A change in the selected model after excluding a study indicated sensitivity to that study, whereas consistent selection across the 5 folds indicated stable model selection.

Candidate scores combined final and process performance. The final-heave component, \(E_{\mathrm{final}}\), was calculated from the absolute percentage errors in final plug heave. The process component, \(E_{\mathrm{curve}}\), measured the error over the complete heave--depth curve and included only studies that reported process histories. To prevent studies with more cases or curves from dominating the score, errors were first averaged within each study and then across studies.

For LOSO scoring, each curve RMSE was normalized by the measured final plug heave:

\begin{equation}
\mathrm{NRMSE}_{\mathrm{curve}}
=
100
\frac{\mathrm{RMSE}_{\mathrm{curve}}}
{h_{\mathrm{meas,final}}}.
\label{eq:loso-curve-nrmse}
\end{equation}

where \(\mathrm{RMSE}_{\mathrm{curve}}\) is the root-mean-square difference between the predicted and measured curves, and \(h_{\mathrm{meas,final}}\) is the measured plug heave at the end of the curve. Averaging these normalized errors within each study and then across the contributing studies gave \(E_{\mathrm{curve}}\).

The primary score assigned 75\% of the weight to final-heave error and 25\% to process-curve error:

\begin{equation}
J
=
0.75E_{\mathrm{final}}
+
0.25E_{\mathrm{curve}}.
\label{eq:loso-primary-score}
\end{equation}

The candidate model with the lowest \(J\) was selected in each fold. The analysis was repeated using final-heave and process-curve weights of 50:50 and 25:75 to assess whether model selection depended on the relative weighting of the two types of evidence.

%=============================================================================
\section*{Data availability}

The published experimental data and model predictions used in the benchmark
validation are provided as Supplementary Data 1. The model inputs and their
sources are provided as Supplementary Data 2.

\section*{Code availability}

The code used to run the final soil-plug solver and reproduce the benchmark predictions is publicly available under the MIT License at https://github.com/shijie774302906/soil-plug-solver.

\bibliography{ref_multi_agent}

\section*{Funding}

J.S. and Z.O. acknowledge support from the National Natural Science
Foundation of China (grant nos. 52401333 and U24B20114).

\section*{Author contributions}

J.S. conducted the research and wrote the original draft.
J.S. and Y.L. prepared the figures. Y.L. revised the manuscript.
Z.O. supervised the research and acquired funding.

\section*{Ethics declaration}
Not applicable.

\section*{Competing interests}

The authors declare no competing interests.

\newpage

%=============================================================================
\clearpage
% Supplementary material input file generated from ../supplementary_arxiv_source/main.tex.
% This file intentionally has no documentclass, package imports, or document environment.
% It inherits the LaTeX template from main.tex.

% =========================================================
% Supplementary Information
% =========================================================
\section*{Supplementary Information}
\label{supp:supplementary-information}

\noindent\textit{Human-guided physics-constrained AI agents construct an
auditable model of soil-plug evolution}
\par\medskip

This Supplementary Information first presents the model equations and numerical
implementation of the soil-plug solver. It then reports the procedures and
findings of the two workflow audits, followed by the checks applied to the
final code. The final note and accompanying data files document the benchmark
inputs, their sources, and the numerical results cited in the main text.

% ---------------------------------------------------------
% Contents
% ---------------------------------------------------------
\subsection*{Contents}
\label{supp:contents}

\textbf{Supplementary Table 1. Contents of the Supplementary Information.}

\begin{table}[H]
\centering
\scriptsize
\begin{tabularx}{\textwidth}{>{\raggedright\arraybackslash}X >{\raggedright\arraybackslash}X}
\toprule
Section
 & 
Content
 \\
\midrule
Supplementary Note 1 & Model formulation and numerical solution procedure. \\
Supplementary Note 2 & Design and findings of the two workflow audits. \\
Supplementary Note 3 & Verification of the final solver through module-level, regression, and analytical reference tests. \\
Supplementary Note 4 & Benchmark inputs, data sources, and comparison results. \\
Supplementary Data 1 & Numerical data underlying the benchmark summaries and Fig. 5 of the main text. \\
Supplementary Data 2 & Sources and uses of the global and case-specific model inputs. \\
\bottomrule
\end{tabularx}
\end{table}

The benchmark evaluation reported in the main text uses the boundary rules,
global settings, and study-specific inputs documented in Supplementary
Tables 10--12. Supplementary Tables 13--16 report the three corresponding
comparisons shown in Fig. 5.

\clearpage

% =========================================================
% Supplementary Note 1. Mathematical Formulation and Numerical Implementation
% =========================================================
\section*{Supplementary Note 1. Mathematical Formulation and Numerical Implementation of the Soil-Plug Model}
\label{supp:note-1-mathematical-formulation-and-numerical-implementation}

Supplementary Note 1 defines how the soil-plug formulation selected in the
main text calculates heave during caisson penetration and how this calculation
is implemented numerically. The formulation begins with solid-volume
conservation, which combines geometric displacement, seepage-induced changes
in soil volume, and near-wall dilation. The following sections derive these
contributions, assemble the final heave calculation, and then present the
discrete solution process, numerical checks, and range of application.

% ---------------------------------------------------------
% S1.1 Overall Model Formulation and Notation
% ---------------------------------------------------------
\subsection*{S1.1 Overall Formulation and Solution Target}
\label{supp:s1-1-overall-model-formulation-and-notation}

The soil-plug model calculates heave by combining three physical contributions.
$M_G$ represents the geometric displacement caused by skirt penetration, $M_S$
represents the change in soil void ratio driven by seepage and effective-stress
changes, and $M_D$ represents near-wall soil dilation. At penetration depth
\(z\), solid-volume conservation combines these contributions as follows:

\begin{equation*}
h_{\mathrm{heave}}(z)
=
\int_0^z
\left[
\alpha_A
\frac{1+e(\zeta_0,z)}{1+e_0}
-1
\right]
\,\mathrm{d}\zeta_0 ,
\tag{S1}
\label{eq:supp-master-plug-heave}
\end{equation*}

where \(\zeta_0\) labels a soil layer by its initial depth below the seabed,
\(e(\zeta_0,z)\) is its void ratio after the $M_S$ update and subsequent $M_D$
correction, and \(e_0\) is the initial void ratio. The area ratio is
\(\alpha_A=(D_o/D_i)^2\), where \(D_o\) and \(D_i\) are the outer and inner
caisson diameters.

Equation~\eqref{eq:supp-master-plug-heave} calculates plug heave from the
void-ratio profile inside the caisson. $M_G$ first establishes
the geometric limit in which the void ratio remains unchanged. To determine
the departure from that limit, $M_S$ converts the prescribed suction-pressure
history into a hydraulic gradient, an effective-stress path, and then a
seepage-driven void-ratio update. After the $M_G$--$M_S$ calculation has
converged, $M_D$ adds the near-wall dilation mobilized during the penetration
increment. The final void-ratio field is then returned to
Eq.~\eqref{eq:supp-master-plug-heave} to calculate plug length and plug heave.

\begingroup

\scriptsize
\setlength{\tabcolsep}{3.5pt}
\renewcommand{\arraystretch}{1.18}
\setlength{\LTleft}{0pt}
% Absorb sub-point width rounding at the right edge of the longtable.
\setlength{\LTright}{\fill}

\noindent
\textbf{Supplementary Table 2. Symbols, parameters, and state conventions used
in the soil-plug formulation.}

\smallskip

\begin{longtable}{
@{}
>{\raggedright\arraybackslash}p{0.28\textwidth}
>{\raggedright\arraybackslash}p{0.54\textwidth}
>{\raggedright\arraybackslash}p{0.14\textwidth}
@{}
}
\toprule
Symbol or mark & Definition & Unit or type \\
\midrule
\endfirsthead
\multicolumn{3}{@{}l}{\textit{Supplementary Table 2 continued}}\\[2pt]
\toprule
Symbol or mark & Definition & Unit or type \\
\midrule
\endhead
\midrule
\multicolumn{3}{r@{}}{\textit{Continued on next page}}\\
\endfoot
\bottomrule
\endlastfoot

\multicolumn{3}{@{}l}{\textit{Geometry and coordinates}} \\
\addlinespace[2pt]
\(D_i,D_o,r_i\) & \(D_i\): inner caisson diameter; \(D_o\): outer caisson diameter; \(r_i\): inner caisson radius. & m \\
\(A_i,\alpha_A\) & \(A_i\): internal cross-sectional area; \(\alpha_A\): outer-to-inner area ratio, \((D_o/D_i)^2\). & \(A_i\): m\(^2\); \(\alpha_A\): dimensionless \\
\(z,z_n,z_{n+1}\) & \(z\): penetration depth; \(z_n\): depth at the start of the current step; \(z_{n+1}\): depth at the end of the current step. & m \\
\(\Delta z,\dot z,\Delta t\) & \(\Delta z\): penetration increment; \(\dot z\): penetration rate; \(\Delta t\): elapsed time for one step. & \(\Delta z\): m; \(\dot z\): m s\(^{-1}\); \(\Delta t\): s \\
\(h_{\mathrm{heave}},L_{\mathrm{plug}},H\) & \(h_{\mathrm{heave}}\): net plug heave; \(L_{\mathrm{plug}}\): total plug length; \(H\): domain height measured from the plug free surface, \(H=L_{\mathrm{plug}}=z+h_{\mathrm{heave}}\). & m \\
\(y,x\) & \(y\): fixed vertical coordinate measured downward from the initial seabed; \(x\): vertical coordinate measured downward from the current plug free surface. & m \\
\midrule
\multicolumn{3}{@{}l}{\textit{Hydraulic parameters and flow rates}} \\
\addlinespace[2pt]
\(\Delta u\) & Prescribed suction-pressure difference. & kPa \\
\(\gamma_w,\gamma'\) & \(\gamma_w\): unit weight of water; \(\gamma'\): buoyant unit weight of sand. & kN m\(^{-3}\) \\
\(\phi\) & Equivalent head loss. & m \\
\(i,i_{\mathrm{tip}}\) & \(i\): hydraulic gradient, \(d\phi/dx\); \(i_{\mathrm{tip}}\): hydraulic gradient at the caisson tip. & dimensionless \\
\(k_v,k_r\) & \(k_v\): equivalent vertical permeability coefficient; \(k_r\): equivalent radial permeability coefficient. & m s\(^{-1}\) \\
\(k_i,k_{\mathrm{out}}\) & \(k_i\): internal-soil permeability; \(k_{\mathrm{out}}\): external-soil permeability. & m s\(^{-1}\) \\
\(R_e^*,R_{e,\infty},R_b\) & \(R_e^*\): adopted equivalent outer-domain radius; \(R_{e,\infty}\): modal radius estimate for an unbounded outer domain; \(R_b\): source-based finite limiting radius. & m \\
\(\ell_s,\lambda_1\) & \(\ell_s\): characteristic seepage length; \(\lambda_1\): first vertical modal wavenumber. & \(\ell_s\): m; \(\lambda_1\): m\(^{-1}\) \\
\(\mathrm{K}_0(\cdot),\mathrm{K}_1(\cdot)\) & \(\mathrm{K}_0(\cdot)\): zeroth-order modified Bessel function of the second kind; \(\mathrm{K}_1(\cdot)\): first-order modified Bessel function of the second kind. & functions \\
\(Q_x,q_r\) & \(Q_x\): signed axial Darcy discharge; \(q_r\): radial seepage rate per unit plug length. & \(Q_x\): m\(^3\) s\(^{-1}\); \(q_r\): m\(^2\) s\(^{-1}\) \\
\(Q_{\mathrm{in,top}},Q_{\mathrm{tip}}\) & \(Q_{\mathrm{in,top}}\): upward flow rate across the plug top; \(Q_{\mathrm{tip}}\): upward flow rate across the caisson-tip section. & m\(^3\) s\(^{-1}\) \\
\(Q_{\mathrm{geom}},Q_{\mathrm{pump}}\) & \(Q_{\mathrm{geom}}\): geometric volume-change rate; \(Q_{\mathrm{pump}}\): equivalent pump flow. & m\(^3\) s\(^{-1}\) \\
\midrule
\multicolumn{3}{@{}l}{\textit{Stress and soil-state parameters}} \\
\addlinespace[2pt]
\(\sigma'_v,\sigma'_h,\sigma'_{rc}\) & \(\sigma'_v\): vertical effective stress; \(\sigma'_h\): lateral effective stress; \(\sigma'_{rc}\): interfacial reference confining stress. & kPa \\
\(\hat{\sigma}'_v,\sigma'_{v,\mathrm{reg}}\) & \(\hat{\sigma}'_v\): regularized vertical effective stress in the refreshed final output; \(\sigma'_{v,\mathrm{reg}}\): regularized vertical effective stress in trial or history states. & kPa \\
\(p',\hat p',q\) & \(p'\): mean effective stress; \(\hat p'\): regularized mean effective stress; \(q\): deviatoric stress magnitude. & kPa \\
\(\eta\) & Stress ratio, \(q/\hat p'\). & dimensionless \\
\(\nu,K_0,K_p\) & \(\nu\): Poisson's ratio; \(K_0\): at-rest earth-pressure coefficient; \(K_p\): passive stress-ratio bound. & dimensionless \\
\(c_p(K_0)\) & Coefficient converting vertical effective stress to the stored mean-stress reference, \((1+2K_0)/3\). & dimensionless \\
\(p'_{\min}\) & Positive numerical stress floor. & kPa \\
\(\phi_{\mathrm{model}}\) & Case-specific friction-angle input. & degrees \\
\(e,e_0\) & \(e\): current void ratio; \(e_0\): initial void ratio. & dimensionless \\
\(e_{\min},e_{\max}\) & \(e_{\min}\): source-based lower void-ratio bound; \(e_{\max}\): source-based upper void-ratio bound. & dimensionless \\
\(e_{cs},\Psi\) & \(e_{cs}\): critical-state void ratio; \(\Psi\): state parameter, \(e-e_{cs}\). & dimensionless \\
\(e_\Gamma,\lambda_c,\kappa_c\) & \(e_\Gamma\): critical-state-line intercept; \(\lambda_c\): critical-state-line shape coefficient; \(\kappa_c\): critical-state-line shape exponent. & dimensionless \\
\(p_a\) & Reference stress in the critical-state relation. & kPa \\
\(M_{tc},M_{te},M_{\mathrm{path}}\) & \(M_{tc}\): triaxial-compression critical stress ratio; \(M_{te}\): triaxial-extension critical stress ratio; \(M_{\mathrm{path}}\): critical stress ratio adopted for the current path. & dimensionless \\
\(\eta_{\mathrm{act}},\eta_0\) & \(\eta_{\mathrm{act}}\): stress ratio at node activation; \(\eta_0\): stored activation reference stress ratio. & dimensionless \\
\(\mu,\bar\mu,\Delta\bar\mu\) & \(\mu\): current path-mobilization degree; \(\bar\mu\): maximum mobilization reached in accepted steps; \(\Delta\bar\mu\): newly mobilized increment. & dimensionless \\
\(\epsilon_\eta\) & Numerical denominator floor in the mobilization calculation. & dimensionless \\
\(\kappa_s\) & Swelling index. & dimensionless \\
\(\Delta e_{\mathrm{reb}},e^{\mathrm{reb}}\) & \(\Delta e_{\mathrm{reb}}\): void-ratio increment due to rebound; \(e^{\mathrm{reb}}\): void ratio after rebound. & dimensionless \\
\(C^{\mathrm{reb}},\Delta e_{\mathrm{rel}}\) & \(C^{\mathrm{reb}}\): remaining dense-state capacity for void-ratio increase after rebound; \(\Delta e_{\mathrm{rel}}\): void-ratio increment released by path mobilization. & dimensionless \\
\midrule
\multicolumn{3}{@{}l}{\textit{Near-wall dilation parameters}} \\
\addlinespace[2pt]
\(D_r,Q\) & \(D_r\): relative density; \(Q\): parameter in the Bolton-type dilation relation. & dimensionless \\
\(c_\psi,p_{a,B}\) & \(c_\psi\): coefficient converting relative dilatancy index to dilatancy angle; \(p_{a,B}\): reference pressure in the dilation relation. & \(c_\psi\): degrees per unit \(I_R\); \(p_{a,B}\): kPa \\
\(I_R,\psi_d\) & \(I_R\): Bolton-type relative dilatancy index; \(\psi_d\): local dilatancy angle. & \(I_R\): dimensionless; \(\psi_d\): rad \\
\(\delta_y,\delta_{\mathrm{cum}}\) & \(\delta_y\): characteristic displacement; \(\delta_{\mathrm{cum}}\): cumulative interfacial displacement recorded in the dilation history. & m \\
\(\Delta\delta_{\mathrm{pot}},\Delta\delta_{\mathrm{mob}}\) & \(\Delta\delta_{\mathrm{pot}}\): potential interfacial displacement increment; \(\Delta\delta_{\mathrm{mob}}\): increment recorded when a positive dilation increment is applied. & m \\
\(\Delta\bar\varepsilon_{v,\max}^{\mathrm D},\allowbreak\Delta\bar\varepsilon_{v,\mathrm{raw}}^{\mathrm D}\) & \(\Delta\bar\varepsilon_{v,\max}^{\mathrm D}\): maximum cross-sectionally averaged dilation strain increment; \(\Delta\bar\varepsilon_{v,\mathrm{raw}}^{\mathrm D}\): increment before capacity truncation. & dimensionless \\
\(\Delta\bar\varepsilon_{v,\mathrm{rem}}^{\mathrm D},\allowbreak\Delta\bar\varepsilon_v^{\mathrm D}\) & \(\Delta\bar\varepsilon_{v,\mathrm{rem}}^{\mathrm D}\): remaining cross-sectionally averaged dilation strain increment; \(\Delta\bar\varepsilon_v^{\mathrm D}\): applied increment. & dimensionless \\
\(\varepsilon_{\mathrm{cum}}^{\mathrm D},\omega\) & \(\varepsilon_{\mathrm{cum}}^{\mathrm D}\): accumulated dilation strain amplitude; \(\omega\): dilation-capacity decay factor. & dimensionless \\
\(\Delta e^{\mathrm D}\) & Void-ratio increment due to dilation. & dimensionless \\
\midrule
\multicolumn{3}{@{}l}{\textit{Discretization, iteration, and state notation}} \\
\addlinespace[2pt]
\(\zeta_{0,j},\Delta\zeta_{0,j}\) & \(\zeta_{0,j}\): initial depth of material node \(j\); \(\Delta\zeta_{0,j}\): trapezoidal quadrature weight assigned to that node. & m \\
\(j,N\) & \(j\): material-node index; \(N\): total number of material nodes. & integers \\
\(n,k,k_{\mathrm{conv}}\) & \(n\): penetration-step index; \(k\): fixed-point iteration index; \(k_{\mathrm{conv}}\): accepted converged iteration index. & integers \\
\(h_{\mathrm{est}},h_{\mathrm{cand}}\) & \(h_{\mathrm{est}}\): plug-heave estimate entering an iteration; \(h_{\mathrm{cand}}\): candidate plug heave calculated in that iteration. & m \\
\(r_h\) & Absolute plug-heave residual used to test fixed-point convergence. & m \\
\(\mathcal I_j^{n+1},\mathcal A_{n+1}\) & \(\mathcal I_j^{n+1}\): node-activation flag for the current step; \(\mathcal A_{n+1}\): active-node set for the current step. & \(\mathcal I_j^{n+1}\): 0 or 1; \(\mathcal A_{n+1}\): set \\
\(\mathcal S_{j,n+1}^{*}\) & Synchronized state of node \(j\) after coupled convergence and before dilation correction. & state vector \\
\(\mathcal F_{n+1}\) & Fixed-point map from trial void ratio and plug heave to their candidate values. & map \\
\(\mathrm{tr},\mathrm H\) & \(\mathrm{tr}\): trial-state mark; \(\mathrm H\): stored-history mark. & superscripts \\
\(\mathrm D,*\) & \(\mathrm D\): dilation-stage mark; \(*\): synchronized-state mark after coupled convergence and before dilation correction. & superscripts \\
\(\mathrm{out},\mathrm{conv}\) & \(\mathrm{out}\): refreshed final-output mark; \(\mathrm{conv}\): converged-iteration mark. & superscripts or labels \\
\(\operatorname{proj},\operatorname{sat},\operatorname{clip}\) & \(\operatorname{proj}\): projection to stress bounds; \(\operatorname{sat}\): restriction to \([0,1]\); \(\operatorname{clip}\): restriction to an admissible interval. & operators \\
\midrule
\multicolumn{3}{@{}l}{\textit{Model components}} \\
\addlinespace[2pt]
\(M_G,M_S,M_D\) & \(M_G\): skirt-penetration-induced geometric displacement; \(M_S\): seepage-driven soil void-ratio evolution; \(M_D\): near-wall soil dilation. & model components \\
\end{longtable}

\endgroup

% ---------------------------------------------------------
% S1.2 Coordinate System and Geometric Baseline
% ---------------------------------------------------------
\subsection*{S1.2 Coordinate System and Geometric Baseline}
\label{supp:s1-2-geometric-baseline-and-coordinate-system}

Let \(z\) denote the caisson penetration depth and
\(h_{\mathrm{heave}}(z)\) the net plug heave. The fixed coordinate \(y\) is
measured downward from the initial mudline. The coordinate attached to the
moving plug free surface is

\begin{equation*}
x=y+h_{\mathrm{heave}}(z), \qquad H=z+h_{\mathrm{heave}}(z),
\tag{S2}
\label{eq:supp-c-1}
\end{equation*}

where \(x\) is positive downward, \(x=0\) denotes the current plug free
surface, \(x=H\) denotes the caisson tip, and
\(H=L_{\mathrm{plug}}(z)\) is the current total plug length.

When the void ratio remains at its initial value,
\(e(\zeta_0,z)=e_0\), Eq.~\eqref{eq:supp-master-plug-heave} reduces to

\begin{equation*}
h_{\mathrm{heave}}(z)=(\alpha_A-1)z.
\tag{S3}
\label{eq:supp-geometric-baseline}
\end{equation*}

This relation defines the $M_G$ geometric baseline. The $M_S$ and $M_D$
components move the model away from this baseline by changing the void-ratio
field used in the solid-volume-conservation relation.

% ---------------------------------------------------------
% S1.3 Seepage-Driven Void-Ratio Evolution ($M_S$)
% ---------------------------------------------------------
\subsection*{S1.3 Seepage-Driven Void-Ratio Evolution ($M_S$)}
\label{supp:s1-3-seepage-driven-void-ratio-evolution}

Prescribed suction first creates a hydraulic head difference across the soil
plug. This difference determines the hydraulic gradient, which changes the
effective stresses carried by the soil skeleton. The resulting stress path then
determines the change in void ratio. $M_S$ therefore follows the sequence
\(\Delta u\rightarrow i\rightarrow(\sigma'_v,\sigma'_h,p',q)
\rightarrow e^*\), where \(e^*\) is the void ratio accepted after the coupled
$M_G$--$M_S$ calculation converges and then passed to $M_D$.

% ---------------------------------------------------------
% S1.3.1 Hydraulic Boundary Conditions and Seepage Kernel
% ---------------------------------------------------------
\subsubsection*{S1.3.1 From Prescribed Suction to the Hydraulic Field}
\label{supp:s1-3-1-hydraulic-boundary-conditions-and-seepage-kernel}

The first step converts the prescribed suction-pressure difference into the
hydraulic gradient along the plug. The hydraulic calculation resolves vertical
Darcy flow along \(x\) and represents seepage through the surrounding soil as
an equivalent radial discharge per unit plug length. The two flow paths form
the one-dimensional axial equation used below.

Let \(\phi(x)\) denote
the equivalent head loss, expressed as water head, and let \(\Delta u(z)>0\)
denote the prescribed suction-pressure difference. With
\(\gamma_w\) denoting the unit weight of water, the boundary conditions are

\begin{equation*}
\phi(0)=0, \qquad \phi(H)=\frac{\Delta u(z)}{\gamma_w}.
\tag{S4}
\label{eq:supp-c-2}
\end{equation*}

The signed axial discharge \(Q_x\) uses the equivalent vertical permeability
\(k_v\) and the internal cross-sectional area \(A_i\):

\begin{equation*}
Q_x=-k_v A_i\frac{d\phi}{dx}
\tag{S5}
\label{eq:supp-axial-darcy-flux}
\end{equation*}

The radial seepage rate per unit plug length, \(q_r\), uses the equivalent
radial permeability \(k_r\), inner radius \(r_i\), and adopted equivalent
outer-domain radius \(R_e^*\). In this equivalent one-dimensional plug domain,
\(q_r>0\) denotes radial discharge leaving the plug domain:

\begin{equation*}
q_r=\frac{2\pi k_r\phi}{\ln(R_e^*/r_i)}.
\tag{S6}
\label{eq:supp-radial-leakage}
\end{equation*}

Conservation of fluid mass gives \(dQ_x/dx+q_r=0\). With \(A_i=\pi r_i^2\), this balance becomes

\begin{equation*}
\frac{d^2\phi}{dx^2}
-
\frac{2k_r}{k_v r_i^2\ln(R_e^*/r_i)}
\phi
=0.
\tag{S7}
\label{eq:supp-seepage-ode}
\end{equation*}

The radius \(R_e^*\) therefore controls the radial drainage resistance
represented in the one-dimensional model.

Defining \(\ell_s\) as the characteristic seepage length gives the equivalent
one-dimensional finite-domain kernel

\begin{equation*}
\frac{d^2\phi}{dx^2} - \frac{\phi}{\ell_s^2} =0,
\tag{S8}
\label{eq:supp-c-3}
\end{equation*}

with

\begin{equation*}
r_i=\frac{D_i}{2}, \qquad
\ell_s = r_i \left[ \frac{k_v}{2k_r} \ln\left( \frac{R_e^*}{r_i} \right) \right]^{1/2},
\tag{S9}
\label{eq:supp-c-4}
\end{equation*}

where \(D_i\) is the inner diameter. For each benchmark case, we set the
equivalent ratio \(k_r/k_v\) in Eq.~\eqref{eq:supp-c-4} equal to the adopted
internal-to-external permeability ratio \(k_i/k_{\mathrm{out}}\) as a model
assumption.

The unbounded modal estimate \(R_{e,\infty}\) is
defined by matching the logarithmic radial drainage resistance of the primary
leakage mode,

\begin{equation*}
\ln\left(\frac{R_{e,\infty}}{r_i}\right)
=
\frac{\mathrm{K}_0(\lambda_1 r_i)}
{\lambda_1 r_i \mathrm{K}_1(\lambda_1 r_i)}, \qquad
\lambda_1=\frac{\pi}{H},
\tag{S10}
\label{eq:supp-unbounded-radius-matching}
\end{equation*}

where \(\lambda_1\) is the first vertical modal wavenumber and \(\mathrm{K}_0(\cdot)\)
and \(\mathrm{K}_1(\cdot)\) are modified Bessel functions of the second kind. When a
finite external boundary or source-based limiting radius \(R_b\) is available,
the adopted radius is \(R_e^*=\min(R_b,R_{e,\infty})\), with \(R_e^*>r_i\).
If no finite limiting radius is available, \(R_e^*=R_{e,\infty}\). The limiting
radius is specified from source boundary information or from the benchmark
parameter rule.

Solving Eq.~\eqref{eq:supp-c-3} with Eq.~\eqref{eq:supp-c-2} gives

\begin{equation*}
\phi(x) = \frac{\Delta u(z)}{\gamma_w} \frac{ \sinh(x/\ell_s) }{ \sinh(H/\ell_s) }.
\tag{S11}
\label{eq:supp-c-5}
\end{equation*}

In the deep-plug limit, and away from the immediate free-surface boundary, Eq.~\eqref{eq:supp-c-5} approaches

\begin{equation*}
\phi(x)
\approx
\frac{\Delta u(z)}{\gamma_w}
\exp\left(\frac{x-H}{\ell_s}\right).
\tag{S12}
\label{eq:supp-seepage-asymptotic-profile}
\end{equation*}

The hydraulic gradient is \(i=d\phi/dx\); positive \(i\) represents the upward
seepage contribution that reduces vertical effective stress:

\begin{equation*}
i(x,z) = \frac{ \Delta u(z) }{ \gamma_w \ell_s \sinh(H/\ell_s) } \cosh(x/\ell_s).
\tag{S13}
\label{eq:supp-c-6}
\end{equation*}

The kernel satisfies the imposed pressure-difference constraint:

\begin{equation*}
\gamma_w \int_0^H i(x,z)\,dx = \Delta u(z).
\tag{S14}
\label{eq:supp-c-7}
\end{equation*}

The hydraulic gradient in Eq.~\eqref{eq:supp-c-6} supplies the seepage-force
term required to calculate the vertical effective stress.

% ---------------------------------------------------------
% S1.3.2 Moving-Boundary Effective Stress and Tip Gradient
% ---------------------------------------------------------
\subsubsection*{S1.3.2 From Hydraulic Gradient to Effective Stress}
\label{supp:s1-3-2-moving-boundary-effective-stress-and-tip-gradient}

Upward seepage reduces the vertical effective stress carried by the soil
skeleton. Combining the submerged soil weight with the seepage-force term gives
the vertical effective stress at a position \(x\) below the current plug free
surface:

\begin{equation*}
\sigma'_v(x,z) = \int_0^x \left[ \gamma' - \gamma_w i(s,z) \right]ds,
\tag{S15}
\label{eq:supp-c-8}
\end{equation*}

where \(\gamma'\) is the buoyant unit weight of sand. The symbol \(s\) is a
temporary variable used to sum the soil-weight and seepage-force contributions
from the plug surface to position \(x\). Substituting
Eq.~\eqref{eq:supp-c-6} gives

\begin{equation*}
\sigma'_v(x,z) = \gamma' x - \Delta u(z) \frac{ \sinh(x/\ell_s) }{ \sinh(H/\ell_s) }.
\tag{S16}
\label{eq:supp-c-9}
\end{equation*}

The free-surface boundary condition is

\begin{equation*}
\sigma'_v(0,z)=0.
\tag{S17}
\label{eq:supp-c-10}
\end{equation*}

The hydraulic gradient at the caisson tip is obtained by evaluating
Eq.~\eqref{eq:supp-c-6} at \(x=H\):

\begin{equation*}
i_{\mathrm{tip}}(z) = i(H,z) = \frac{ \Delta u(z) }{ \gamma_w \ell_s } \coth(H/\ell_s).
\tag{S18}
\label{eq:supp-c-11}
\end{equation*}

The resulting vertical stress field is then used to calculate the lateral,
mean, and deviatoric stresses. The corresponding stress path controls the
void-ratio update.

% ---------------------------------------------------------
% S1.3.3 Stress-Path Mobilization and Void-Ratio Update
% ---------------------------------------------------------
\subsubsection*{S1.3.3 From Effective-Stress Path to Void-Ratio Change}
\label{supp:s1-3-3-stress-path-mobilization-and-void-ratio-update}

At each penetration depth, we use the effective stresses to calculate changes
in void ratio and obtain plug heave from Eq.~\eqref{eq:supp-master-plug-heave}.
The effective stresses depend on seepage through the plug, which is itself
affected by the plug height. Updating the void ratio thus changes the height
used to calculate seepage and effective stresses. We account for this dependence
by repeatedly recalculating seepage and stresses with the updated plug geometry,
then updating the void ratio and heave. This procedure is implemented as a
fixed-point iteration.

Within each iteration, the void-ratio update combines two components.
Changes in mean effective stress produce a rebound component, with unloading
increasing the void ratio. A second component allows a further increase toward
the critical-state void ratio as the stress ratio approaches its limiting value.
For penetration step
\(n+1\), \(k\) identifies the current iteration. The trial geometry defines the
total plug length \(H_{n+1}^{(k)}\) and the free-surface-attached coordinate
\(x_{j,n+1}^{(k)}\) of node \(j\). The corresponding vertical effective stress is

\begin{equation*}
\sigma_{v,j,n+1}^{\prime (k)} =
\gamma' x_{j,n+1}^{(k)}
- \Delta u(z_{n+1})
\frac{\sinh\left(x_{j,n+1}^{(k)}/\ell_s^{(k)}\right)}
{\sinh\left(H_{n+1}^{(k)}/\ell_s^{(k)}\right)} .
\tag{S19}
\label{eq:supp-d-3}
\end{equation*}

Nodes with \(\sigma_{v,j,n+1}^{\prime (k)}\le 0\) are flagged as near-critical
seepage locations. To keep the logarithmic and stress-ratio calculations well
defined, the current trial stress and the stress stored from the preceding
accepted step are bounded below by \(p'_{\min}\):

\begin{equation*}
\begin{aligned}
\sigma_{v,j,n+1,\mathrm{reg}}^{\prime (k)}
&=
\max \left[ \sigma_{v,j,n+1}^{\prime (k)}, p'_{\min} \right],\\
\sigma_{v,j,n,\mathrm{reg}}^{\prime \mathrm{H}}
&=
\max \left[ \sigma_{v,j,n}^{\prime \mathrm{H}}, p'_{\min} \right],
\end{aligned}
\tag{S20}
\label{eq:supp-d-4}
\end{equation*}
where \(p'_{\min}\) is a positive numerical stress floor and the superscript
\(\mathrm H\) denotes a history value stored from the preceding accepted step.

The lateral effective stress is first estimated from the change in regularized
vertical stress relative to the preceding accepted state:

\begin{equation*}
\sigma_{h,j,n+1}^{\prime \mathrm{tr},(k)} =
\sigma_{h,j,n}^{\prime \mathrm{H}}
+
\frac{\nu}{1-\nu}
\left(
\sigma_{v,j,n+1,\mathrm{reg}}^{\prime (k)}
-
\sigma_{v,j,n,\mathrm{reg}}^{\prime \mathrm{H}}
\right),
\tag{S21}
\label{eq:supp-d-5}
\end{equation*}
where \(\nu\) is Poisson's ratio. The trial value is then limited between the
numerical stress floor and the passive stress bound:

\begin{equation*}
\sigma_{h,j,n+1}^{\prime (k)}
=
\operatorname{proj}
\left[
\sigma_{h,j,n+1}^{\prime \mathrm{tr},(k)};
p'_{\min},
K_p\sigma_{v,j,n+1,\mathrm{reg}}^{\prime (k)}
\right].
\tag{S22}
\label{eq:supp-d-6}
\end{equation*}

The projection operator and passive stress-ratio bound are defined as

\begin{equation*}
\operatorname{proj}[a;l,u]=\min[\max(a,l),u],
\qquad
K_p = \frac{1+\sin\phi_{\mathrm{model}}}
{1-\sin\phi_{\mathrm{model}}} ,
\tag{S23}
\label{eq:supp-d-7}
\end{equation*}

where \(a\) is the value being projected and \(l\) and \(u\) are its lower and
upper bounds. The quantity \(K_p\) is the passive stress-ratio bound, and
\(\phi_{\mathrm{model}}\) is the case-specific friction-angle input used
by the equivalent one-dimensional solver.

The vertical and lateral effective stresses are used to calculate the mean
effective stress and the stress ratio. Of these, the mean effective stress is
used to determine the rebound response and the critical-state void ratio,
while the stress ratio is
used to quantify progress toward the limiting stress state.

The mean effective stress \(p'\) and the corresponding regularized value
\(\hat p'\) are

\begin{equation*}
p_{j,n+1}^{\prime (k)}
=
\frac{ \sigma_{v,j,n+1,\mathrm{reg}}^{\prime (k)} + 2\sigma_{h,j,n+1}^{\prime (k)} }{3},
\qquad
\hat p_{j,n+1}^{\prime (k)}
=
\max \left[ p_{j,n+1}^{\prime (k)}, p'_{\min} \right].
\tag{S24}
\label{eq:supp-d-8}
\end{equation*}

The deviatoric stress magnitude \(q\) and stress ratio \(\eta=q/\hat p'\) are

\begin{equation*}
q_{j,n+1}^{(k)}
=
\left|
\sigma_{v,j,n+1,\mathrm{reg}}^{\prime (k)}
-
\sigma_{h,j,n+1}^{\prime (k)}
\right|,
\qquad
\eta_{j,n+1}^{(k)}
=
\frac{q_{j,n+1}^{(k)}}{\hat p_{j,n+1}^{\prime (k)}}.
\tag{S25}
\label{eq:supp-d-9}
\end{equation*}

The critical-state void ratio provides the reference for the additional
void-ratio increase. Its value at the current mean effective stress is

\begin{equation*}
e_{cs,j,n+1}^{(k)}
=
e_\Gamma
-
\lambda_c
\left(
\frac{\hat p_{j,n+1}^{\prime (k)}}{p_a}
\right)^{\kappa_c} ,
\tag{S26}
\label{eq:supp-d-10}
\end{equation*}

where \(e_\Gamma\), \(\lambda_c\), and \(\kappa_c\) are the intercept and shape
parameters of the adopted critical-state line, and \(p_a\) is its reference
stress.

% ---------------------------------------------------------
To measure progress toward the limiting stress state, the current stress ratio
is compared with its value when the soil first enters the caisson and with the
applicable stress-ratio limit. The entry value is stored for each node as

\begin{equation*}
\eta_{0,j}=\eta_{j,\mathrm{act}} .
\tag{S27}
\label{eq:supp-d-11}
\end{equation*}

The friction-angle input defines separate stress-ratio limits for triaxial
compression and extension:

\begin{equation*}
M_{tc}
=
\frac{6\sin\phi_{\mathrm{model}}}
{3-\sin\phi_{\mathrm{model}}},
\qquad
M_{te}
=
\frac{6\sin\phi_{\mathrm{model}}}
{3+\sin\phi_{\mathrm{model}}}.
\tag{S28}
\label{eq:supp-triaxial-stress-ratios}
\end{equation*}

The applicable limit depends on whether the vertical or lateral effective
stress is larger:

\begin{equation*}
M_{\mathrm{path},j,n+1}^{(k)} =
\begin{cases}
  M_{tc}, & \sigma_{v,j,n+1,\mathrm{reg}}^{\prime (k)} \ge \sigma_{h,j,n+1}^{\prime (k)}, \\[4pt]
  M_{te}, & \sigma_{h,j,n+1}^{\prime (k)} > \sigma_{v,j,n+1,\mathrm{reg}}^{\prime (k)} .
\end{cases}
\tag{S29}
\label{eq:supp-d-12}
\end{equation*}

The mobilization degree \(\mu\) quantifies the increase in stress ratio from
the entry value \(\eta_{0,j}\) toward the selected limit. The operator
\(\operatorname{sat}[a,0,1]=\min[\max(a,0),1]\) bounds its argument \(a\)
between 0 and 1:

\begin{equation*}
\mu_{j,n+1}^{(k)}
=
\operatorname{sat}
\left[
\frac{
\eta_{j,n+1}^{(k)}-\eta_{0,j}
}{
\max\left(
M_{\mathrm{path},j,n+1}^{(k)}-\eta_{0,j},
\epsilon_{\eta}
\right)
},
0,1
\right],
\tag{S30}
\label{eq:supp-d-13}
\end{equation*}
where \(\epsilon_{\eta}\) is a positive lower bound used to keep the
denominator nonzero.

Values of \(\mu=0\), \(0<\mu<1\), and \(\mu=1\) represent no, partial,
and full mobilization, respectively.

The rebound contribution is then calculated from the change in mean effective
stress relative to the preceding accepted step. It uses the swelling index
\(\kappa_s\) and the positive mean-stress reference
\(p_{j,n}^{\prime \mathrm H}\) stored at that step:

\begin{equation*}
\Delta e_{\mathrm{reb},j}^{(k)}
=
-
\kappa_s
\ln
\left(
\frac{\hat p_{j,n+1}^{\prime (k)}}{p_{j,n}^{\prime \mathrm{H}}}
\right),
\qquad
e_{j,n+1}^{\mathrm{reb},(k)}
=
e_{j,n}
+
\Delta e_{\mathrm{reb},j}^{(k)} .
\tag{S31}
\label{eq:supp-d-14}
\end{equation*}

After the rebound update, the remaining capacity for an additional void-ratio
increase is the positive gap between the critical-state void ratio and the
value after rebound:

\begin{equation*}
C_{j,n+1}^{\mathrm{reb},(k)}
=
\max
\left[
e_{cs,j,n+1}^{(k)}
-
e_{j,n+1}^{\mathrm{reb},(k)},
0
\right].
\tag{S32}
\label{eq:supp-d-15}
\end{equation*}

To count each mobilization increase once, the current degree is compared with
the maximum saved through the preceding accepted step. The positive difference
gives the new mobilization increment:

\begin{equation*}
\Delta \bar{\mu}_{j,n+1}^{(k)}
=
\max
\left[
\mu_{j,n+1}^{(k)}
-
\bar{\mu}_{j,n},
0
\right],
\tag{S33}
\label{eq:supp-d-16}
\end{equation*}
where \(\bar{\mu}_{j,n}\) is the maximum mobilization degree committed through
step \(n\).

Multiplying this new mobilization increment by the remaining capacity gives
the additional void-ratio increase:

\begin{equation*}
\Delta e_{\mathrm{rel},j}^{(k)}
=
\Delta \bar{\mu}_{j,n+1}^{(k)}
C_{j,n+1}^{\mathrm{reb},(k)} .
\tag{S34}
\label{eq:supp-d-17}
\end{equation*}

Adding this increase to the void ratio after rebound gives the trial void
ratio. The operator
\(\operatorname{clip}[a,l,u]=\min[\max(a,l),u]\) limits this value to the
source-based admissible bounds \(e_{\min}\) and \(e_{\max}\):

\begin{equation*}
e_{j,n+1}^{*,(k)}
=
\operatorname{clip}
\left[
e_{j,n+1}^{\mathrm{reb},(k)}
+
\Delta e_{\mathrm{rel},j}^{(k)},
e_{\min},
e_{\max}
\right].
\tag{S35}
\label{eq:supp-d-18}
\end{equation*}

The superscript \(*\) denotes the coupled-loop state before the \(M_D\)
correction. The state parameter \(\Psi=e-e_{cs}\) measures the difference
between this updated void ratio and its critical-state value:

\begin{equation*}
\Psi_{j,n+1}^{*,(k)}
=
e_{j,n+1}^{*,(k)}
-
e_{cs,j,n+1}^{(k)} .
\tag{S36}
\label{eq:supp-ms-state-parameter}
\end{equation*}

% ---------------------------------------------------------
% S1.4 Near-Wall Soil Dilation ($M_D$)
% ---------------------------------------------------------
\subsection*{S1.4 Near-Wall Soil Dilation ($M_D$)}
\label{supp:s1-4-near-wall-soil-dilation}

$M_D$ calculates the additional void-ratio increment caused by near-wall soil
dilation after the $M_G$--$M_S$ calculation has converged. To obtain this
increment, $M_D$ needs the local dilation tendency, the interface displacement
mobilized during the current penetration step, and the dilation capacity
remaining in the current soil state. These quantities jointly determine the
void-ratio increment $\Delta e_j^{\mathrm D}$, which is added to the converged
void ratio.

% ---------------------------------------------------------
% S1.4.1 Converged Input State and Near-Wall Confinement
% ---------------------------------------------------------
\subsubsection*{S1.4.1 Converged Input State and Near-Wall Confinement}
\label{supp:s1-4-1-dilation-tendency-from-converged-state}

To calculate the local dilation tendency, $M_D$ establishes the normal
confinement acting near the caisson wall. For each active soil layer, the
calculation begins from the void ratio and effective stresses returned by the
converged $M_G$--$M_S$ calculation:

\begin{equation*}
\left\{
e_j^{*},\,
\sigma_{v,j}^{\prime *},\,
\sigma_{h,j}^{\prime *}
\right\},
\tag{S37}
\label{eq:supp-e-2}
\end{equation*}

where the superscript \(*\) marks the synchronized state accepted at
$M_G$--$M_S$ convergence. The lateral effective stress provides the wall
confinement used in the dilation relation. The reference confinement is
bounded below by \(p'_{\min}\) as a numerical safeguard:

\begin{equation*}
\sigma'_{rc,j}
=
\max
\left(
\sigma_{h,j}^{\prime *},
p'_{\min}
\right).
\tag{S38}
\label{eq:supp-e-3}
\end{equation*}

% ---------------------------------------------------------
% S1.4.2 Dilation Tendency and Potential Interface Displacement
% ---------------------------------------------------------
\subsubsection*{S1.4.2 Dilation Tendency and Potential Interface Displacement}
\label{supp:s1-4-2-dilation-tendency-and-mobilized-displacement}

With the wall confinement established, $M_D$ calculates whether the soil tends
to dilate and the strength of that tendency. The adopted Bolton-type relation
expresses this tendency as the relative dilatancy index
\(I_{R,j}\)~\cite{bolton1986strength}. The relation evaluates the index from
the relative density \(D_r\), material parameter \(Q\), reference pressure
\(p_{a,B}\), and wall confinement:

\begin{equation*}
I_{R,j}
=
\max
\left[
D_r
\left(
Q-\ln\frac{\max(\sigma'_{rc,j},p_{a,B})}{p_{a,B}}
\right)
-1,
0
\right],
\tag{S39}
\label{eq:supp-e-4}
\end{equation*}

where \(D_r\) is expressed as a decimal. When the calculated confinement is
below \(p_{a,B}\), this lower bound sets the logarithmic term to zero.
The angle coefficient \(c_\psi\) converts \(I_{R,j}\) into the
local dilatancy angle:

\begin{equation*}
\psi_{d,j}
=
\frac{\pi}{180}
c_{\psi}I_{R,j}.
\tag{S40}
\label{eq:supp-e-5}
\end{equation*}

The factor \(\pi/180\) converts the degree-valued intermediate
\(c_\psi I_{R,j}\) to radians. A positive \(I_{R,j}\) therefore produces a
positive dilation tendency, whereas \(I_{R,j}=0\) gives
\(\psi_{d,j}=0\).

The dilation angle describes the tendency to expand, but the amount developed
during one penetration step also depends on the interface displacement
mobilized in that step. The current penetration increment is
\(\Delta z=z_{n+1}-z_n\). The parameter \(\delta_y\) denotes the characteristic
mobilization displacement, and \(\delta_{\mathrm{cum},j}^{n}\) is the cumulative
interfacial displacement committed through the preceding step.
Equation~\eqref{eq:supp-e-6} combines these quantities to give the potential
displacement for the current step:

\begin{equation*}
\Delta\delta_{\mathrm{pot},j}
=
\delta_y
\left[
1-\exp
\left(
-\frac{\Delta z}{\delta_y}
\right)
\right]
\exp
\left(
-\frac{\delta_{\mathrm{cum},j}^{n}}{\delta_y}
\right).
\tag{S41}
\label{eq:supp-e-6}
\end{equation*}

The bracketed term describes displacement mobilization over the current
penetration increment. The final exponential factor reduces further
mobilization as previously committed interface displacement accumulates.
For \(\Delta z\ge 0\), the resulting
\(\Delta\delta_{\mathrm{pot},j}\) is non-negative.

% ---------------------------------------------------------
% S1.4.3 State Capacity and Void-Ratio Correction
% ---------------------------------------------------------
\subsubsection*{S1.4.3 State Capacity and Void-Ratio Correction}
\label{supp:s1-4-3-state-capacity-and-void-ratio-correction}

To determine the limit on further dilation, $M_D$ compares the current void
ratio with the critical-state value. This critical-state value is calculated
from the mean effective stress in the converged $M_G$--$M_S$ state. When the
current void ratio is lower, the positive difference is used to calculate the
maximum dilation strain:

\begin{equation*}
\begin{aligned}
p_j^{\prime \mathrm{D},*}
&=
\max\left[
\frac{\sigma_{v,j}^{\prime *}+2\sigma_{h,j}^{\prime *}}{3},
p'_{\min}
\right],\\
e_{cs,j}^{\mathrm{D},*}
&=
e_\Gamma-
\lambda_c\left(\frac{p_j^{\prime \mathrm{D},*}}{p_a}\right)^{\kappa_c},\\
\Psi_j^{\mathrm{D},*}
&=
e_j^*-e_{cs,j}^{\mathrm{D},*},\\
\Delta\bar{\varepsilon}_{v,\max,j}^{\mathrm{D}}
&=
\frac{\max(-\Psi_j^{\mathrm{D},*},0)}{1+e_j^{*}}.
\end{aligned}
\tag{S42}
\label{eq:supp-e-7}
\end{equation*}

where \(\Psi_j^{\mathrm{D},*}<0\) indicates that the current void ratio is
below the critical-state value. Dividing this positive void-ratio gap by
\(1+e_j^*\) gives the maximum cross-sectionally averaged dilation strain
\(\Delta\bar{\varepsilon}_{v,\max,j}^{\mathrm D}\).

The current increment also depends on two distinct histories. The cumulative
interface displacement
\(\delta_{\mathrm{cum},j}^{n}\) in Eq.~\eqref{eq:supp-e-6} controls further
displacement mobilization. The cumulative dilation-stage strain amplitude
\(\varepsilon_{\mathrm{cum},j}^{\mathrm D,n}\) records the dilation already
applied through step \(n\). The factor \(\omega_j\) reduces the current
increment as this accumulated dilation approaches the current capacity:

\begin{equation*}
\omega_j =
\begin{cases}
\max \left[ 0, 1- \dfrac{ \varepsilon_{\mathrm{cum},j}^{\mathrm{D},n} }{ \Delta\bar{\varepsilon}_{v,\max,j}^{\mathrm{D}} } \right], & \Delta\bar{\varepsilon}_{v,\max,j}^{\mathrm{D}}>0 \ \text{and}\ I_{R,j}>0, \\[12pt]
0, & \Delta\bar{\varepsilon}_{v,\max,j}^{\mathrm{D}}\le 0 \ \text{or}\ I_{R,j}=0 .
\end{cases}
\tag{S43}
\label{eq:supp-e-8}
\end{equation*}

The local wall expansion is converted into an equivalent strain averaged
over the caisson cross-section. The product
\(\Delta\delta_{\mathrm{pot},j}\sin\psi_{d,j}\) gives the outward displacement
normal to the wall, and \(4/D_i\) is the perimeter-to-area ratio of the internal
caisson section. Multiplying by \(\omega_j\) gives the strain after history
attenuation, before the remaining-capacity limit is applied:

\begin{equation*}
\Delta\bar{\varepsilon}_{v,\mathrm{raw},j}^{\mathrm{D}}
=
\frac{
4\Delta\delta_{\mathrm{pot},j}\sin\psi_{d,j}
}{
D_i
}
\omega_j .
\tag{S44}
\label{eq:supp-e-9}
\end{equation*}

The remaining capacity is obtained by subtracting the dilation already
applied from the maximum amplitude:

\begin{equation*}
\Delta\bar{\varepsilon}_{v,\mathrm{rem},j}^{\mathrm{D}}
=
\max
\left[
\Delta\bar{\varepsilon}_{v,\max,j}^{\mathrm{D}}
-
\varepsilon_{\mathrm{cum},j}^{\mathrm{D},n},
0
\right].
\tag{S45}
\label{eq:supp-e-10}
\end{equation*}

The applied amplitude is the smaller of the untruncated value and this
remaining capacity. When this amplitude is positive, the full potential
interface displacement is recorded as \(\Delta\delta_{\mathrm{mob},j}\);
otherwise, the recorded displacement is zero:

\begin{equation*}
\begin{aligned}
\Delta\bar{\varepsilon}_{v,j}^{\mathrm{D}}
&=
\min
\left[
\Delta\bar{\varepsilon}_{v,\mathrm{raw},j}^{\mathrm{D}},
\Delta\bar{\varepsilon}_{v,\mathrm{rem},j}^{\mathrm{D}}
\right],\\
\Delta\delta_{\mathrm{mob},j}
&=
\begin{cases}
\Delta\delta_{\mathrm{pot},j}, & \Delta\bar{\varepsilon}_{v,j}^{\mathrm{D}}>0,\\[2pt]
0, & \Delta\bar{\varepsilon}_{v,j}^{\mathrm{D}}=0.
\end{cases}.
\end{aligned}
\tag{S46}
\label{eq:supp-e-11}
\end{equation*}

The applied dilation-stage strain amplitude gives the void-ratio increment:

\begin{equation*}
\Delta e_j^{\mathrm{D}}
=
(1+e_j^{*})
\Delta\bar{\varepsilon}_{v,j}^{\mathrm{D}}.
\tag{S47}
\label{eq:supp-e-12}
\end{equation*}

Adding this increment to the converged $M_S$ void ratio and limiting the result
to the admissible bounds gives the corrected void ratio for the current
penetration step:

\begin{equation*}
e_j^{n+1}
=
\operatorname{clip}
\left[
e_j^{*}
+
\Delta e_j^{\mathrm{D}},
e_{\min},
e_{\max}
\right].
\tag{S48}
\label{eq:supp-e-13}
\end{equation*}

The applied dilation-stage amplitude also updates the cumulative history
carried into the subsequent penetration step:

\begin{equation*}
\varepsilon_{\mathrm{cum},j}^{\mathrm{D},n+1}
=
\varepsilon_{\mathrm{cum},j}^{\mathrm{D},n}
+
\Delta\bar{\varepsilon}_{v,j}^{\mathrm{D}}.
\tag{S49}
\label{eq:supp-e-14}
\end{equation*}

% ---------------------------------------------------------
% S1.5 Integrated Plug-Heave Assembly and Output Quantities
% ---------------------------------------------------------
\subsection*{S1.5 Integrated Plug-Heave Assembly and Output Quantities}
\label{supp:s1-5-integrated-plug-heave-assembly-and-output-quantities}

The final void-ratio field returned by $M_D$ is used to calculate plug length
and heave by solid-volume conservation. The resulting geometry provides the
domain for the refreshed hydraulic, stress, and pump-flow outputs.

% ---------------------------------------------------------
% S1.5.1 Final Geometry, Output Refresh, and State Commitment
% ---------------------------------------------------------
\subsubsection*{S1.5.1 Final Geometry, Output Refresh, and State Commitment}
\label{supp:s1-5-1-final-geometry-output-refresh-and-history-submission}

The final geometry is assembled from the material layers reached by the
caisson at \(z_{n+1}\). For material node \(j\), the active-node flag
\(\mathcal{I}_{j}^{n+1}\) equals 1 when
\(\zeta_{0,j}\leq z_{n+1}\) and 0 otherwise. The active set collects the nodes
with \(\mathcal{I}_{j}^{n+1}=1\):

\begin{equation*}
\mathcal{A}_{n+1}
=
\left\{
j
\mid
\mathcal{I}_{j}^{n+1}=1
\right\}.
\tag{S50}
\label{eq:supp-e-15}
\end{equation*}

For each active material interval, solid-volume conservation converts its
initial thickness \(\Delta\zeta_{0,j}\) into its current contribution to plug
length. The ratio \((1+e_j^{n+1})/(1+e_0)\) accounts for the change in bulk soil
volume at fixed solid volume. The area ratio \(\alpha_A\) maps the reference
layer volume associated with the outer caisson cross-section to the internal
cross-section. Summing these contributions gives the plug length:

\begin{equation*}
L_{\mathrm{plug}}^{n+1}
=
\sum_{j\in \mathcal{A}_{n+1}}
\alpha_A
\frac{ 1+e_j^{n+1} }{ 1+e_0 }
\Delta \zeta_{0,j},
\tag{S51}
\label{eq:supp-e-16}
\end{equation*}

where \(\Delta\zeta_{0,j}\) is the trapezoidal quadrature weight assigned to
node \(j\) over the active material interval.

The assembled plug length contains the penetrated portion below the seabed and
the heaved portion above it. Subtracting the current penetration depth gives
the net plug heave:

\begin{equation*}
h_{\mathrm{heave}}^{n+1}
=
L_{\mathrm{plug}}^{n+1}
-
z_{n+1}.
\tag{S52}
\label{eq:supp-e-17}
\end{equation*}

Equations~\eqref{eq:supp-e-16} and \eqref{eq:supp-e-17} are the cellwise
discretization of Eq.~\eqref{eq:supp-master-plug-heave}.

The corrected geometry also sets the domain height used to refresh the
reported hydraulic and stress fields. The output-state height is reconstructed
from the penetration depth and plug heave:

\begin{equation*}
H_{n+1}^{\mathrm{out}}
=
z_{n+1}
+
h_{\mathrm{heave}}^{n+1}.
\tag{S53}
\label{eq:supp-e-18}
\end{equation*}

The resulting \(H_{n+1}^{\mathrm{out}}\) equals the assembled plug length. The
superscript \(\mathrm{out}\) marks its use in the output refresh.

The final void-ratio field maps each active material coordinate to its output
coordinate:

\begin{equation*}
x_j^{\mathrm{out}}
=
\int_0^{\zeta_{0,j}}
\alpha_A
\frac{1+e^{n+1}(\zeta_0)}{1+e_0}
\,d\zeta_0 ,
\tag{S54}
\label{eq:supp-output-material-coordinate}
\end{equation*}

where \(e^{n+1}(\zeta_0)\) denotes the final void-ratio field expressed over the
material coordinate.

The output seepage length is the finite-domain characteristic length evaluated with the refreshed plug length,

\begin{equation*}
\ell_s^{\mathrm{out}}
=
r_i
\left[
\frac{k_v}{2k_r}
\ln\left(
\frac{R_e^{*,\mathrm{out}}}{r_i}
\right)
\right]^{1/2},
\tag{S55}
\label{eq:supp-output-seepage-length}
\end{equation*}
where \(R_e^{*,\mathrm{out}}\) follows the adopted outer-domain rule in Section
S1.3.1, evaluated with \(H=H_{n+1}^{\mathrm{out}}\). The refreshed hydraulic and vertical
stress fields are evaluated from the same finite-domain relations:

\begin{equation*}
i_j^{\mathrm{out}}
=
\frac{ \Delta u(z_{n+1}) }
{ \gamma_w \ell_s^{\mathrm{out}}
\sinh \left( H_{n+1}^{\mathrm{out}}/\ell_s^{\mathrm{out}} \right) }
\cosh
\left(
\frac{x_j^{\mathrm{out}}}{\ell_s^{\mathrm{out}}}
\right),
\tag{S56}
\label{eq:supp-e-19}
\end{equation*}

\begin{equation*}
\sigma_{v,j}^{\prime \mathrm{out}}
=
\gamma' x_j^{\mathrm{out}}
-
\Delta u(z_{n+1})
\frac{
\sinh
\left(
x_j^{\mathrm{out}}/\ell_s^{\mathrm{out}}
\right)
}{
\sinh
\left(
H_{n+1}^{\mathrm{out}}/\ell_s^{\mathrm{out}}
\right)
}.
\tag{S57}
\label{eq:supp-e-20}
\end{equation*}

The refreshed vertical stress is first bounded below by \(p'_{\min}\). The
converged lateral stress is then projected onto the bounds associated with this
refreshed vertical stress:

\begin{equation*}
\hat\sigma_{v,j}^{\prime \mathrm{out}}
=
\max
\left[
\sigma_{v,j}^{\prime \mathrm{out}},
p'_{\min}
\right],
\qquad
\sigma_{h,j}^{\prime \mathrm{out}}
=
\operatorname{proj}
\left[
\sigma_{h,j}^{\prime *};
p'_{\min},
K_p\hat\sigma_{v,j}^{\prime \mathrm{out}}
\right].
\tag{S58}
\label{eq:supp-output-stress-projection}
\end{equation*}

The refreshed mean and deviatoric stresses stored for output are

\begin{equation*}
\begin{aligned}
p_j^{\prime \mathrm{out}}
&=
\frac{
\hat\sigma_{v,j}^{\prime \mathrm{out}}
+
2\sigma_{h,j}^{\prime \mathrm{out}}
}{3},\\
\hat p_j^{\prime \mathrm{out}}
&=
\max
\left[
p_j^{\prime \mathrm{out}},
p'_{\min}
\right],\\
q_j^{\mathrm{out}}
&=
\left|
\hat\sigma_{v,j}^{\prime \mathrm{out}}
-
\sigma_{h,j}^{\prime \mathrm{out}}
\right|.
\end{aligned}
\tag{S59}
\label{eq:supp-e-21}
\end{equation*}

After the $M_G$--$M_S$ iteration has ended, the solver evaluates this refresh
once from the final corrected geometry and stores the resulting hydraulic and
stress fields as outputs.

At the end of the step, the final void ratio and applied interfacial
displacement are committed:

\begin{equation*}
e_{j,n+1}=e_j^{n+1},
\qquad
\delta_{\mathrm{cum},j}^{n+1}
=
\delta_{\mathrm{cum},j}^{n}
+
\Delta\delta_{\mathrm{mob},j}.
\tag{S60}
\label{eq:supp-e-22}
\end{equation*}

% ---------------------------------------------------------
% S1.5.2 Equivalent Pump-Flow Post-Processing
% ---------------------------------------------------------
\subsubsection*{S1.5.2 Equivalent Pump-Flow Post-Processing}
\label{supp:s1-5-2-pump-flow-post-processing}

After the final geometry and hydraulic field have been refreshed, the model
calculates the equivalent pump flow associated with the prescribed suction.
This quantity combines the Darcy inflow across the plug top with the rate of
change in internal plug volume. The hydraulic quantities used below are taken
from the refreshed output state: \(H=H_{n+1}^{\mathrm{out}}\),
\(\ell_s=\ell_s^{\mathrm{out}}\), and \(i=i^{\mathrm{out}}\).

The internal cross-sectional area is

\begin{equation*}
A_i=\frac{\pi D_i^2}{4}.
\tag{S61}
\label{eq:supp-c-13}
\end{equation*}

Using the refreshed hydraulic field, we calculate the upward flow rates
across the plug top, \(Q_{\mathrm{in,top}}\), and the caisson tip,
\(Q_{\mathrm{tip}}\). Both are defined as positive upward:

\begin{equation*}
Q_{\mathrm{in,top}}
=
-Q_x(0)
=
A_i k_v i(0,z)
=
A_i k_v
\frac{\Delta u(z)}
{\gamma_w \ell_s \sinh(H/\ell_s)},
\qquad
Q_{\mathrm{tip}} = -Q_x(H)= A_i k_v i_{\mathrm{tip}}.
\tag{S62}
\label{eq:supp-c-12}
\end{equation*}

The second contribution is the change in internal plug volume during the
penetration step. Because \(H=z+h_{\mathrm{heave}}\), this change contains both
the penetration increment and the heave increment. For successive accepted
steps \(n\) and \(n+1\), let \(\Delta z=z_{n+1}-z_n\), let \(\dot z\) denote the
penetration rate, and let \(\Delta t\) denote the elapsed time:

\begin{equation*}
Q_{\mathrm{geom}}
=
A_i
\left(
\frac{z_{n+1}-z_n}{\Delta t}
+
\frac{h_{\mathrm{heave}}^{n+1}-h_{\mathrm{heave}}^n}{\Delta t}
\right),
\qquad
\Delta t=\frac{\Delta z}{\dot z}.
\tag{S63}
\label{eq:supp-c-14}
\end{equation*}

The equivalent pump-flow estimate is the sum of the top inflow and the
geometric volume-change rate:

\begin{equation*}
Q_{\mathrm{pump}} = Q_{\mathrm{in,top}} + Q_{\mathrm{geom}}.
\tag{S64}
\label{eq:supp-c-15}
\end{equation*}

The solver records \(Q_{\mathrm{pump}}\) after the output refresh. Calculating
the physical flow rate requires a measured time interval or penetration rate.
For benchmark profiles without timing data, the calculation uses a unit time
interval, \(\Delta t=1\); the resulting pump-flow values are excluded from
the plug-heave benchmark assessment.

\clearpage

% ---------------------------------------------------------
% S1.6 Discrete Solution and State Transfer
% ---------------------------------------------------------
\subsection*{S1.6 Discrete Solution and State Transfer}
\label{supp:s1-6-discrete-solution-and-state-submission}

The numerical solver advances through the prescribed penetration depths,
using the accepted state from one step to initialize the next
(Supplementary Fig. 1).

\begin{figure}[H]
\centering
\includegraphics[width=0.9\textwidth]{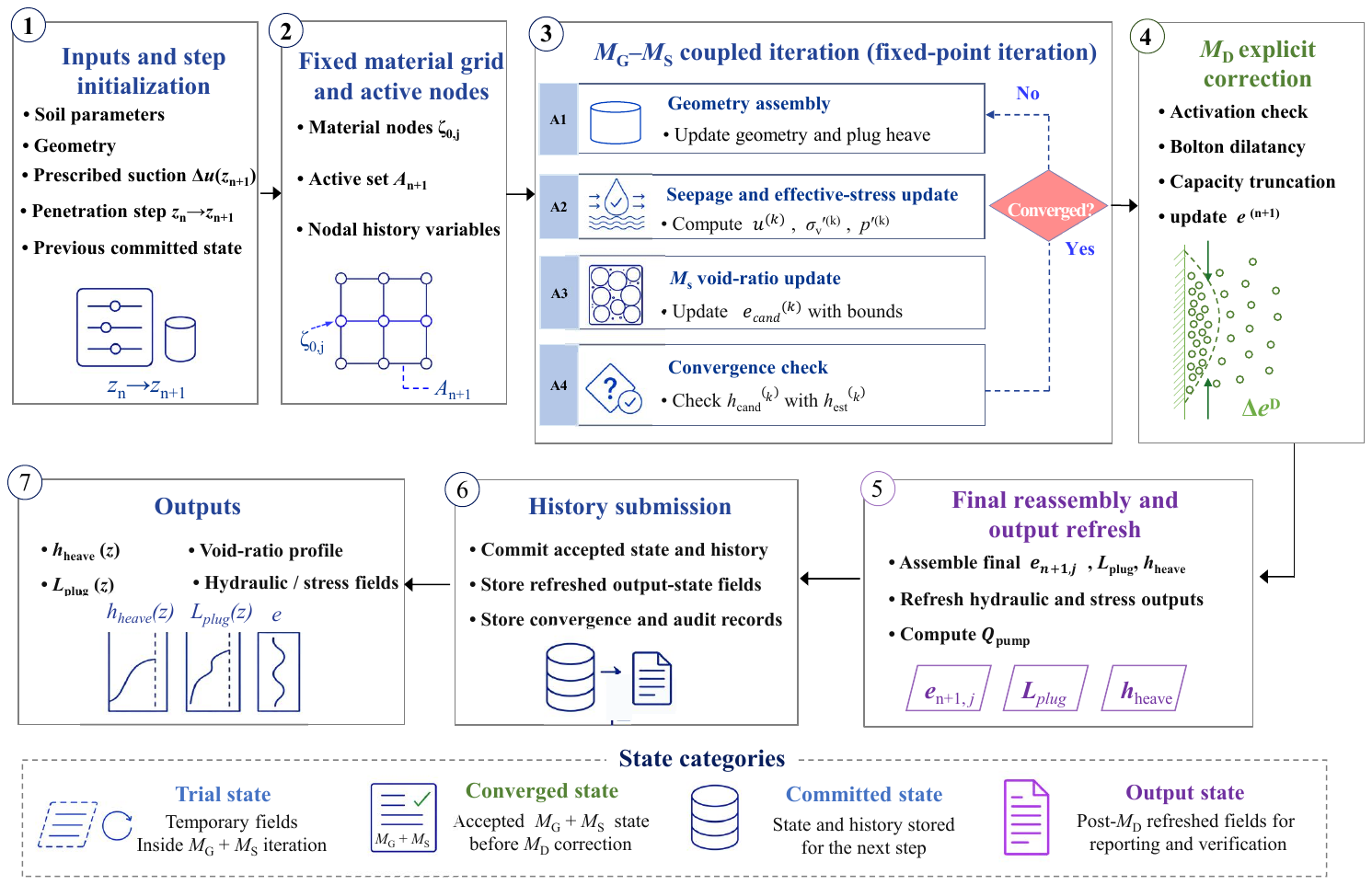}
\caption*{\textbf{Supplementary Fig. 1. Stepwise numerical solution and state
transfer of the soil-plug model.} Trial state denotes fields updated during
the $M_G$--$M_S$ iteration; converged state denotes the accepted solution before
the $M_D$ correction. Committed state carries the accepted state and history
into the next penetration step. Output state contains the post-$M_D$ refreshed
fields stored for reporting and verification.}
\end{figure}

% ---------------------------------------------------------
% S1.6.1 Material Grid and Pre-Step State
% ---------------------------------------------------------
\subsubsection*{S1.6.1 Material Grid and Pre-Step State}
\label{supp:s1-6-1-material-grid-and-submitted-state}

The solver represents the initial soil profile on a fixed material grid. The
coordinate \(\zeta_{0,j}\) is the initial depth of node \(j\) below the seabed,
with \(j=1,\ldots,N\). At each penetration step, the solver applies the
active-node rule to this grid. Active nodes contribute to the plug-heave
integral and receive the $M_S$ and $M_D$ updates. For nodes that were already
active at the accepted step
\(n\), the solver reads the following committed nodal state before calculating
step \(n+1\):

\begin{equation*}
\left\{
e_{j,n},\,
\sigma_{v,j,n}^{\prime \mathrm{H}},\,
\sigma_{h,j,n}^{\prime \mathrm{H}},\,
p_{j,n}^{\prime \mathrm{H}},\,
\eta_{0,j},\,
\bar{\mu}_{j,n},\,
\delta_{\mathrm{cum},j}^{n},\,
\varepsilon_{\mathrm{cum},j}^{\mathrm D,n}
\right\}.
\tag{S65}
\label{eq:supp-d-2}
\end{equation*}

The superscript \(\mathrm H\) marks stress quantities carried from the
preceding accepted step. The void ratio \(e_{j,n}\) is the final value after
the preceding $M_D$ correction. The vertical stress
\(\sigma_{v,j,n}^{\prime \mathrm H}\) is obtained from the refreshed final
geometry, while \(\sigma_{h,j,n}^{\prime \mathrm H}\) is the lateral stress
synchronized at $M_G$--$M_S$ convergence. The rebound calculation uses the
positive reference stress
\(p_{j,n}^{\prime \mathrm H}=\max[c_p(K_0)
\sigma_{v,j,n}^{\prime \mathrm H},p'_{\min}]\), where
\(c_p(K_0)=(1+2K_0)/3\).

The remaining quantities record the constitutive history of each node. The
reference \(\eta_{0,j}\) stores the stress ratio at first activation, and
\(\bar{\mu}_{j,n}\) stores the largest path mobilization reached through step
\(n\). The variables \(\delta_{\mathrm{cum},j}^{n}\) and
\(\varepsilon_{\mathrm{cum},j}^{\mathrm D,n}\) record the applied $M_D$
history. The trial calculations use this committed state as a fixed reference.
After the current step is accepted, its updated values become the pre-step
state for the next depth.

A node that first becomes active at step \(n+1\) starts from the initial void
ratio \(e_0\), its geostatic vertical stress \(\gamma'\zeta_{0,j}\), and the
corresponding \(K_0\) lateral stress. Its accumulated path mobilization,
interfacial displacement, and dilation-stage strain amplitude start from zero.
The activation reference \(\eta_{0,j}\) is recorded from the synchronized state
obtained during this first active step. In the stress relations, \(K_0\) denotes
the at-rest earth-pressure coefficient; \(\mathrm{K}_0(\lambda_1r_i)\) denotes the
modified Bessel function used in the hydraulic formulation.

% ---------------------------------------------------------
Each benchmark profile is represented by 121 uniformly spaced material nodes
extending from the initial seabed to the final penetration depth. The input
depths are sorted and duplicate depths are removed to define the penetration
steps; each retained depth uses its paired suction value. Spatial volume
integrals use trapezoidal quadrature. When a penetration depth falls between
adjacent grid nodes, the terminal value is linearly interpolated and the
remaining partial interval is included in the integral.

% ---------------------------------------------------------
% S1.6.2 Fixed-Point Iteration and Convergence
% ---------------------------------------------------------
\subsubsection*{S1.6.2 Fixed-Point Iteration and Convergence}
\label{supp:s1-6-2-fixed-point-iteration-and-convergence}

Each iteration begins with a plug-heave estimate
\(h_{\mathrm{est}}^{(k)}\) and a trial void-ratio field
\(e_{\mathrm{trial}}^{(k)}\). These values determine the plug geometry,
seepage length, hydraulic field, and effective stresses. The stress update
produces a candidate void-ratio field \(e_{\mathrm{cand}}^{(k)}\), which is
integrated to obtain the candidate height \(h_{\mathrm{cand}}^{(k)}\). With
the committed history from step \(n\) held fixed, this calculation defines the
fixed-point map:

\begin{equation*}
\left(e_{\mathrm{cand}}^{(k)},h_{\mathrm{cand}}^{(k)}\right)
=
\mathcal F_{n+1}\!\left(e_{\mathrm{trial}}^{(k)},h_{\mathrm{est}}^{(k)}\right),
\qquad
\left(e^*,h_{\mathrm{heave}}^*\right)
=
\mathcal F_{n+1}\!\left(e^*,h_{\mathrm{heave}}^*\right).
\tag{S66}
\label{eq:supp-fixed-point-map}
\end{equation*}

At a fixed point, the input estimate and returned candidate describe the same
coupled geometry and void-ratio state. The solver measures this closure through
the difference between the height used to construct the trial fields and the
height returned by the updated void ratio:

\begin{equation*}
r_h^{(k)}=
\left|
h_{\mathrm{cand}}^{(k)}-
h_{\mathrm{est}}^{(k)}
\right| .
\tag{S67}
\label{eq:supp-d-19}
\end{equation*}

The iteration is accepted when either the relative residual is below
\(10^{-5}\) or the absolute residual is below \(10^{-8}\,\mathrm{m}\):

\begin{equation*}
\frac{r_h^{(k)}}
{\max\left(\left|h_{\mathrm{est}}^{(k)}\right|,
10^{-10}\,\mathrm{m}\right)}
<10^{-5}
\quad\text{or}\quad
r_h^{(k)}<10^{-8}\,\mathrm{m}.
\tag{S68}
\label{eq:supp-d-20}
\end{equation*}

The relative criterion scales the change by the current heave estimate. The
absolute criterion provides a direct tolerance when the heave is close to zero,
and the \(10^{-10}\,\mathrm{m}\) denominator bound prevents division by zero.

When the criterion is not satisfied, the next estimate is
\(h_{\mathrm{est}}^{(k+1)}=h_{\mathrm{est}}^{(k)}+
0.7[h_{\mathrm{cand}}^{(k)}-h_{\mathrm{est}}^{(k)}]\), and the new void-ratio
candidate becomes the trial field for the next iteration. This partial update
limits oscillation in the feedback between geometry and soil state. Each
penetration step is allowed at most 120 iterations. If no iteration satisfies
Eq.~\eqref{eq:supp-d-20}, the solver reports non-convergence and ends the step
before state transfer.

After the criterion is satisfied, the solver performs one synchronization pass
at the accepted candidate geometry. This pass recalculates the $M_G$--$M_S$
fields so that the returned void ratio, stresses, and plug heave refer to the
same geometry. The accepted iteration is denoted by \(k_{\mathrm{conv}}\). Its
mobilization degree \(\mu_{j,n+1}^{\mathrm{conv}}\) updates the historical
maximum at node \(j\):

\begin{equation*}
\bar{\mu}_{j,n+1}
=
\max
\left[
\bar{\mu}_{j,n},
\mu_{j,n+1}^{\mathrm{conv}}
\right].
\tag{S69}
\label{eq:supp-d-21}
\end{equation*}

The synchronized void-ratio field becomes the pre-dilation value passed to
$M_D$:

\begin{equation*}
e_{j,n+1}^{*}
=
e_{j,n+1}^{*,(k_{\mathrm{conv}})} .
\tag{S70}
\label{eq:supp-d-22}
\end{equation*}

The same synchronization pass returns the corresponding vertical stress,
lateral stress, mean effective stress, critical-state void ratio, and state
parameter:

\begin{equation*}
\begin{aligned}
\sigma_{v,j,n+1}^{\prime *}
&=
\sigma_{v,j,n+1}^{\prime (k_{\mathrm{conv}})},
\qquad
\sigma_{h,j,n+1}^{\prime *}
=
\sigma_{h,j,n+1}^{\prime (k_{\mathrm{conv}})},\\
p_{j,n+1}^{\prime *}
&=
p_{j,n+1}^{\prime (k_{\mathrm{conv}})},
\qquad
e_{cs,j,n+1}^{*}
&=
e_{cs,j,n+1}^{(k_{\mathrm{conv}})},
\qquad
\Psi_{j,n+1}^{*}
&=
\Psi_{j,n+1}^{*,(k_{\mathrm{conv}})} .
\end{aligned}
\tag{S71}
\label{eq:supp-converged-ms-state}
\end{equation*}

These quantities form the synchronized pre-dilation state:

\begin{equation*}
\mathcal{S}_{j,n+1}^{*}
=
\left\{
e_{j,n+1}^{*},
\sigma_{v,j,n+1}^{\prime *},
\sigma_{h,j,n+1}^{\prime *},
p_{j,n+1}^{\prime *},
e_{cs,j,n+1}^{*},
\Psi_{j,n+1}^{*},
\bar{\mu}_{j,n+1}
\right\}.
\tag{S72}
\label{eq:supp-d-23}
\end{equation*}

The star marks quantities evaluated in this synchronized pre-dilation state.
$M_D$ acts once on this state, after which the corrected void ratio determines
the final geometry and output refresh. The accepted histories are then
committed for the next penetration depth.

% ---------------------------------------------------------
% S1.7 Contribution and Numerical Treatment of $M_D$
% ---------------------------------------------------------
\subsection*{S1.7 Contribution and Update-Order Sensitivity of $M_D$}
\label{supp:s1-7-influence-of-the-explicit-md-correction}

To evaluate the influence of $M_D$ after $M_G$--$M_S$ convergence, we
performed two numerical checks across the 14 benchmark profiles. One compared
predictions from $M_G$ alone, $M_G$--$M_S$, and the complete
$M_G$--$M_S$--$M_D$ calculation to quantify the contribution of $M_D$. The
other tested whether one $M_D$ update per penetration step was sufficient by
comparing it with repeated updates at the same penetration depth.

In the reported calculation, each penetration step completed $M_G$--$M_S$
convergence before $M_D$ corrected the void-ratio field and plug heave. After
this single correction, the calculation advanced to the next penetration
depth. Thus, ``once'' means one $M_D$ correction per penetration step. Because
this correction also changes the plug geometry used by $M_G$--$M_S$, the
comparison calculation held the penetration depth fixed and repeated the
update. The $M_D$-corrected heave and void ratio were returned to
$M_G$--$M_S$, after which $M_D$ was applied again. Repetition stopped when two
successive post-$M_D$ heaves differed by less than $10^{-5}$ m, with a maximum
of eight repetitions. The nine final-state profiles comprised four Kim and Kim
cases (T1-1 S 25g, T1-2 S 50g, T3-1 S 50g, and T4-3 S 25g) and five Tran and
Randolph cases (TRAN-T1--TRAN-T5). The five continuous-process profiles
comprised Tjelta TJELTA-A3, Wang full process, and Zhao ZHAO-T1--ZHAO-T3.

\textbf{Supplementary Table 3. Contribution and update-order sensitivity of $M_D$
across the 14 benchmark profiles.}

\begin{table}[H]
\centering
\scriptsize
\begin{tabularx}{\textwidth}{>{\raggedright\arraybackslash}X >{\raggedright\arraybackslash}X >{\raggedright\arraybackslash}X}
\toprule
Check
 & 
Metric
 & 
Value
 \\
\midrule
Component combination & Mean absolute final-heave error with $M_G$ only & 58.42\% \\
Component combination & Mean absolute final-heave error with $M_G$ and $M_S$ & 9.01\% \\
Component combination & Mean absolute final-heave error with $M_G$, $M_S$, and $M_D$ & 9.48\% \\
Relative $M_D$ heave increment & Mean absolute ratio of the $M_D$ heave increment to final plug heave, all profiles & 0.956\% \\
Relative $M_D$ heave increment & Mean absolute ratio of the $M_D$ heave increment to final plug heave, final state prediction & 0.097\% \\
Relative $M_D$ heave increment & Mean absolute ratio of the $M_D$ heave increment to final plug heave, continuous process prediction & 2.503\% \\
Relative $M_D$ heave increment & Maximum absolute ratio of the $M_D$ heave increment to final plug heave & 3.478\% (Zhao T3) \\
Absolute $M_D$ heave increment & Mean absolute $M_D$ heave increment, all profiles & 0.0675 cm \\
Absolute $M_D$ heave increment & Mean absolute $M_D$ heave increment, final state prediction & 0.0475 cm \\
Absolute $M_D$ heave increment & Mean absolute $M_D$ heave increment, continuous process prediction & 0.1035 cm \\
Single versus repeated $M_D$ update & Mean absolute final-heave difference, all profiles & 0.0091 cm \\
Single versus repeated $M_D$ update & Maximum absolute final-heave difference & 0.0200 cm (Wang full process) \\
Single versus repeated $M_D$ update & Maximum absolute difference between final-heave errors & 0.418\% (Wang full process) \\
\bottomrule
\end{tabularx}
\end{table}

Adding $M_S$ to $M_G$ reduced the mean absolute final-heave error from 58.42\%
to 9.01\%, representing the dominant change in the component comparison.
Including $M_D$ changed the aggregate error to 9.48\%, while its mean absolute
contribution was 0.0675 cm, or 0.956\% of the final heave. This contribution
was more visible in the continuous-process profiles than in the final-state
profiles, averaging 2.503\% and 0.097\%, respectively, and reaching 3.478\% for
Zhao T3. These results identify $M_D$ as a modest, case-dependent correction to
the predicted heave; the principal reduction in final-heave error is associated
with $M_S$.

Repeating the $M_D$ update at the same penetration depth changed the final
heave by 0.0091 cm on average and by at most 0.0200 cm. The largest absolute
difference between the two final-heave errors was 0.418\%, also for the Wang
full-process profile. The close agreement across the 14 profiles supports
one $M_D$ correction after $M_G$--$M_S$ convergence in each penetration step.

% ---------------------------------------------------------
% S1.8 Numerical Safeguards and Model Scope
% ---------------------------------------------------------
\subsection*{S1.8 Numerical Safeguards and Model Scope}
\label{supp:s1-8-numerical-safeguards-and-applicability}

Three safeguards keep the calculation well defined near limiting states. The
solver uses \(p'_{\min}=0.1\) kPa as a positive stress floor and
\(\epsilon_{\eta}=10^{-8}\) as the lower bound in the mobilization denominator.
The void-ratio update is limited to the case-specific, source-based bounds
\(e_{\min}\) and \(e_{\max}\), which define the admissible density range. The
bounded stresses enter the constitutive update, but the solver also records the
unbounded vertical stress. A value of \(\sigma'_v\le 0\) indicates that the
assumed seepage field has reduced the computed vertical effective stress to zero
or below.

The hydraulic equations are evaluated only when \(H>0\). At the initial state
\(H=0\), the zero-suction branch assigns zero hydraulic gradients and flow
rates. The seepage model assumes saturated sand and Darcy flow, with
quasi-steady conditions over each penetration step. It uses the equivalent
outer-domain influence radius \(R_e^*\) and holds \(k_v\) and \(k_r\) constant
within that step.

The model does not resolve piping or particle migration. To monitor this
boundary, the solver retains the hydraulic gradient at the caisson tip,
\(i_{\mathrm{tip}}\), at every penetration step. The recorded
\(i_{\mathrm{tip}}\) can be compared with a piping criterion appropriate to the
soil and hydraulic boundary. A step
that reaches or exceeds that criterion lies at the limit of the present
continuum calculation.

The formulation is intended for saturated sand when seepage can be represented
by Darcy flow and the soil--caisson interface response can be represented by a
cross-sectional average. The model would require extension where permeability
changes strongly during installation, deformation localizes into shear bands,
or the interface response becomes strongly three-dimensional. Complex
elastoplastic stress paths and soil-fabric reconstruction after failure or
liquefaction also lie outside its present scope.

\clearpage

% =========================================================
% Supplementary Note 2. Workflow Audit Procedures and Results
% =========================================================
\section*{Supplementary Note 2. Workflow Audit Procedures and Results}
\label{supp:note-2-workflow-execution-and-audit-records}

The main text reports the results of two audits that examined the equations,
code, and check records retained during model development. This note explains
what each audit role received, which answers were withheld, how its response
was scored, and what problems it found.

% ---------------------------------------------------------
% S2.1 Materials Shown to Each Audit Role
% ---------------------------------------------------------
\subsection*{S2.1 Materials Shown to Each Audit Role}
\label{supp:s2-1-audit-materials-and-information-boundaries}

We used two audit tasks to test whether an audit role could find problems
without being given their answers. In the blinded replay, the role received
materials related to a problem previously identified during development, but
not its label or answer key. The audit role then had to identify and locate the
problem from those materials. For each derivation-to-code probe, a code-writing
role first implemented the supplied equations and completed a fixed set of
checks. A separate audit role then examined the equations, generated code, and
check results to identify problems that those checks did not cover.
Supplementary Table 4 specifies the materials shown and withheld in each task.

\textbf{Supplementary Table 4. Materials shown to and withheld from each audit role.}

\begin{table}[H]
\centering
\scriptsize
\begin{tabularx}{\textwidth}{>{\raggedright\arraybackslash}p{0.17\textwidth} >{\raggedright\arraybackslash}X >{\raggedright\arraybackslash}X >{\raggedright\arraybackslash}p{0.22\textwidth}}
\toprule
Audit task
&
Materials shown to the audit role
&
Information not shown
&
Question evaluated
\\
\midrule
Blinded audit replay (A1--A9) &
The relevant physical relation, calculation rule, required input and output of the code module, uncorrected code or output, and available check records &
The issue label, answer key, prior discussion, correction history, and other project files &
Can the audit role find and locate a known issue without being given its answer? \\
Derivation-to-code probes (P1--P5) &
The supplied equations and state-update rules, generated code, and results of the fixed implementation checks &
The target problem and answer &
Can the audit role identify omissions or inconsistencies not covered by the predeclared checks? \\
\bottomrule
\end{tabularx}
\end{table}

% ---------------------------------------------------------
% S2.2 How the Audits Were Scored
% ---------------------------------------------------------
\subsection*{S2.2 How the Audits Were Scored}
\label{supp:s2-2-scoring-and-passing-criteria}

We scored both audit tasks using criteria that linked each reported problem to
a specific equation, calculation step, or code location. For the blinded
replay, we compared each report with an answer key held outside the audit
packet. A replay passed when it identified the target problem and its location,
explained the physical or numerical consequence, and proposed a valid
correction with a follow-up check. We recorded false positives and human
assistance separately.

For each derivation-to-code probe, a report passed when it identified a
specific disagreement between the supplied equations or update rules and the
generated code, located the affected code, and cited the supporting check
result. Supplementary Table 5 gives the complete scoring criteria.

\textbf{Supplementary Table 5. What counted as a pass in each audit.}

\begin{table}[H]
\centering
\scriptsize
\begin{tabularx}{\textwidth}{>{\raggedright\arraybackslash}p{0.23\textwidth} >{\raggedright\arraybackslash}X >{\raggedright\arraybackslash}X}
\toprule
Scored item
&
What was checked
&
Criterion for a pass
\\
\midrule
Finding and location &
Whether the report found the target problem and named the affected equation, calculation step, or code location &
The target problem and its location are both identified correctly \\
Consequence and correction &
Physical or numerical consequence, proposed correction, and follow-up check &
The correction addresses the problem and the follow-up check can verify it \\
False positives and human assistance &
Any unrelated finding and any help given during replay &
The target problem is found without an incorrect additional finding or an extra hint \\
Independent probe finding &
The equation or rule that the code failed to follow, its code location, and the supporting check result &
The report identifies a specific omission or disagreement from the supplied equations, code, and check results \\
\bottomrule
\end{tabularx}
\end{table}

% ---------------------------------------------------------
% S2.3 Blinded Audit-Replay Results
% ---------------------------------------------------------
\subsection*{S2.3 Blinded Audit-Replay Results}
\label{supp:s2-3-blinded-audit-replay-results}

The audit role found and located all nine target problems without false
positives or human assistance. Supplementary Table 6 lists the target problem
in each replay and the check needed to identify it.

\textbf{Supplementary Table 6. Blinded audit-replay results.}

\begin{table}[H]
\centering
\scriptsize
\begin{tabularx}{\textwidth}{>{\raggedright\arraybackslash}p{0.07\textwidth} >{\raggedright\arraybackslash}X >{\raggedright\arraybackslash}p{0.25\textwidth}}
\toprule
Case
&
Target problem
&
What the audit role had to check
\\
\midrule
A1 & Correction for near-wall soil dilation absent from the solver call chain & Whether the solver calls the $M_D$ correction \\
A2 & Degree-valued \(\psi_d\) passed directly to a trigonometric function expecting radians & Whether the angle unit matches the function input \\
A3 & Committed local increment not limited by the remaining dilation capacity & Whether the applied increment remains within the available capacity \\
A4 & The reported mean effective stress and deviatoric stress retained their trial values and were not recalculated from the final stresses & Whether the reported mean effective stress and deviatoric stress are recalculated from the final stresses \\
A5 & Trial iterations overwrite committed history before convergence is accepted & Whether history is saved only after convergence \\
A6 & Reference confinement uses vertical trial stress instead of lateral trial stress & Which stress is used as the reference confinement \\
A7 & Pump-flow post-processing uses the tip-flow component instead of the top inflow component & Which flow component enters the reported pump flow \\
A8 & Final hydraulic state reuses the trial values instead of recomputing from final geometry & Whether the hydraulic field is recalculated from the final geometry \\
A9 & Post-rebound capacity is computed from the pre-rebound void ratio instead of the rebound-updated value & Which void ratio is used to calculate the remaining capacity \\
\bottomrule
\end{tabularx}
\end{table}

For A6, the initial audit packet omitted the stress-regularization relation
needed to determine the reference confinement. We supplied this equation before
scoring. The issue label and answer key remained withheld.

% ---------------------------------------------------------
% S2.4 Derivation-to-Code Probe Results
% ---------------------------------------------------------
\subsection*{S2.4 Derivation-to-Code Probe Results}
\label{supp:s2-4-derivation-to-code-probe-results}

All 36 fixed implementation checks passed across the five probes, yet the
independent audit identified five additional problems in P1--P4
(Supplementary Table 7). The findings involved code-branch handling,
incomplete instructions, an incorrect count of checked trials, and a disagreement
between an equation and its implementation. The audit found no additional problem when
it repeated the final hydraulic and stress checks in P5.

\textbf{Supplementary Table 7. Derivation-to-code probe results.}

\begin{table}[H]
\centering
\scriptsize
\begin{tabularx}{\textwidth}{>{\raggedright\arraybackslash}p{0.07\textwidth} >{\raggedright\arraybackslash}X >{\raggedright\arraybackslash}p{0.13\textwidth} >{\raggedright\arraybackslash}X}
\toprule
Probe
&
Calculation checked
&
Fixed checks passed
&
What the independent audit found
\\
\midrule
P1 & Recalculation of final outputs in converged, edge, and nonconverged branches & 10/10 & One branch omitted the final-output recalculation; an additional check was added \\
P2 & Active nodes and the piecewise \(\omega\) calculation in $M_D$ & 11/11 & One code branch could fail, and the supplied instructions did not state clearly whether inactive grid nodes remained in the input \\
P3 & Saving $M_S$ history only after convergence is accepted & 5/5 &
The code stopped checking trials as soon as it found one that met the
convergence criterion. However, the recorded count included all supplied
trials, including those that had not been checked. \\
P4 & Applying $M_S$ rebound before calculating the remaining capacity & 5/5 & The code and equation treated \(\Delta e_{\mathrm{reb}}\) differently in the numerator \\
P5 & Recalculating hydraulic and stress outputs from the final geometry & 5/5 & All final-refresh checks were repeated and no additional problem was found \\
\bottomrule
\end{tabularx}
\end{table}

% ---------------------------------------------------------
% S2.5 What the Workflow Audits Establish
% ---------------------------------------------------------
\subsection*{S2.5 What the Workflow Audits Establish}
\label{supp:s2-5-evidence-boundary}

Within the evaluated tasks, the replay and probe results support two distinct
audit functions: recovering known problems from retained records and
identifying implementation problems missed by fixed checks. Benchmark
comparisons provide separate evidence for prediction accuracy.

\clearpage

% =========================================================
% Supplementary Note 3. Checks of the Final Code
% =========================================================
\section*{Supplementary Note 3. Checks of the Final Code}
\label{supp:note-3-implementation-verification-and-regression-baselines}

After addressing the audit findings, we checked whether the final
penetration-step solver followed the equations, update order, and
state-commitment rules defined in Supplementary Note 1. Module-level tests
examined individual calculations and state updates. Analytical reference
checks compared selected relations with independent evaluations, whereas
regression tests checked whether unchanged input sequences reproduced the
stored outputs.

% ---------------------------------------------------------
% S3.1 Module-Level Tests
% ---------------------------------------------------------
\subsection*{S3.1 Module-Level Tests}
\label{supp:s3-1-module-level-verification}

All 33 module-level tests passed. Supplementary Table 8 groups the tests by
the part of the calculation examined and reports the main operations,
acceptance criteria, and outcome for each group.

\textbf{Supplementary Table 8. Module-level tests and results.}

\begin{table}[H]
\centering
\scriptsize
\begin{tabularx}{\textwidth}{>{\raggedright\arraybackslash}p{0.23\textwidth} >{\centering\arraybackslash}p{0.09\textwidth} >{\raggedright\arraybackslash}X >{\raggedright\arraybackslash}p{0.25\textwidth} >{\centering\arraybackslash}p{0.08\textwidth}}
\toprule
Test group
&
Number of tests
&
Main operations checked
&
Acceptance criterion and observed result
&
Outcome
\\
\midrule
Hydraulic closure and moving-boundary stress &
10 &
Pressure-difference recovery, tip gradient, free-surface stress, suction-controlled boundaries, and pump-flow separation &
Integral and boundary residuals remain within the adopted tolerance; pump flow does not update plug heave or void ratio &
Passed \\
Bolton relationship and $M_D$ closure &
9 &
Node activation, reference confinement, dilatancy, mobilized displacement, capacity truncation, and final clipping &
The correction is non-negative, capacity-limited, and confined to active nodes &
Passed \\
$M_S$ path update and projection constraints &
6 &
Mean effective stress, lateral projection, critical-state variables, rebound, state-gap release, and clipping &
Updated variables match analytical or independently evaluated values &
Passed \\
State commitment and final output refresh &
8 &
Separation of trial and accepted states, saving after convergence, final $M_D$ update, output recalculation, and saved iteration and check results &
Histories are saved only from accepted states, and final outputs are recalculated from the final geometry and stresses &
Passed \\
\bottomrule
\end{tabularx}
\end{table}

% ---------------------------------------------------------
% S3.2 Analytical Reference and Regression Checks
% ---------------------------------------------------------
\subsection*{S3.2 Analytical Reference and Regression Checks}
\label{supp:s3-2-analytical-reference-and-regression-checks}

All five analytical reference checks and four regression tests passed
(Supplementary Table 9).

\textbf{Supplementary Table 9. Analytical reference and regression checks.}

\begin{table}[H]
\centering
\scriptsize
\begin{tabularx}{\textwidth}{>{\raggedright\arraybackslash}p{0.08\textwidth} >{\raggedright\arraybackslash}p{0.24\textwidth} >{\raggedright\arraybackslash}X >{\raggedright\arraybackslash}p{0.24\textwidth} >{\centering\arraybackslash}p{0.08\textwidth}}
\toprule
ID
&
Check
&
Quantity or comparison
&
Criterion and observed result
&
Outcome
\\
\midrule
R-01 & Pressure-difference closure & \(\gamma_w\int_0^H i(x,z)\,dx-\Delta u(z)\) & Absolute residual remains within the adopted tolerance & Passed \\
R-02 & Moving free-surface stress & \(\sigma'_v(0,z)\) & Zero within numerical tolerance & Passed \\
R-03 & Deep-limit approximation & Finite-domain seepage kernel and its exponential deep-limit form & Relative difference remains within tolerance & Passed \\
R-04 & Critical-state consistency & \(e_{cs}\) and \(\Psi=e-e_{cs}\) & Matches independent scalar evaluation & Passed \\
R-05 & $M_D$ capacity truncation & \(\Delta\bar{\varepsilon}_{v}^{\mathrm{D}}\le \Delta\bar{\varepsilon}_{v,\mathrm{rem}}^{\mathrm{D}}\) & No node exceeds the remaining capacity & Passed \\
B-01 & Input-sequence consistency & Penetration depths, suction-pressure history, geometry, and soil inputs & Input arrays and parameter assignments remain unchanged & Passed \\
B-02 & $M_G$--$M_S$ baseline & \(h_{\mathrm{heave}}^{\mathrm{GS}}\), \(e^*\), and convergence status & Stored main-chain outputs are reproduced & Passed \\
B-03 & $M_D$ contribution baseline & \(\Delta h_{\mathrm{heave}}^{\mathrm{D}}\), \(\Delta e^{\mathrm{D}}\), and capacity flags & Stored correction outputs are reproduced & Passed \\
B-04 & Post-$M_D$ refresh baseline & Final plug heave, refreshed stresses, and pump-flow outputs & Stored final outputs are reproduced & Passed \\
\bottomrule
\end{tabularx}
\end{table}

% ---------------------------------------------------------
% S3.3 Evidence from the Final-Code Checks
% ---------------------------------------------------------
\subsection*{S3.3 Results of the Final-Code Checks}
\label{supp:s3-3-evidence-boundary}

Within the tested scope, these results show that the code follows the declared
equations, update order, and state-commitment rules.

\clearpage

% =========================================================
% Supplementary Note 4. Benchmark Data, Input Sources, and Results
% =========================================================
\section*{Supplementary Note 4. Benchmark Data, Input Sources, and Comparison Results}
\label{supp:note-4-benchmark-data-parameter-sources-and-benchmark-summaries}

% ---------------------------------------------------------
% S4.1 Public Benchmark Files
% ---------------------------------------------------------
\subsection*{S4.1 Public Benchmark Files}
\label{supp:s4-1-benchmark-data-package-scope}

The public files separate the values used in the benchmark comparisons from the
sources and uses of the model inputs. Supplementary Data 1 provides the
case-level values for three benchmark comparisons: formula reduction and volume
closure, final state prediction, and continuous process prediction.
Supplementary Data 2 records each adopted global and case-specific input, its
source and source class, and where it enters the calculation.
Definitions of all fields in the two data files are provided in
\nolinkurl{SUPPLEMENTARY_DATA_README.md}.

Supplementary Data 2 uses five source classes for adopted inputs:
\begin{itemize}
\item \texttt{direct}: a value reported by a source.
\item \texttt{derived}: a value calculated from reported values.
\item \texttt{design\_parameter}: a value adopted from site or design information.
\item \texttt{analog}: a value transferred from a cited related source.
\item \texttt{assumption}: a value fixed by a declared modeling rule.
\end{itemize}
The label \texttt{unavailable} marks a quantity for
which no sourced value was adopted. The label \texttt{mixed} marks a ledger row
that combines adopted values from different source classes; the source-basis
field identifies the class of each value.

% ---------------------------------------------------------
% S4.2 Shared Inputs
% ---------------------------------------------------------
\subsection*{S4.2 Shared Input Rules}
\label{supp:s4-3-shared-uncertain-inputs}

The final-state and continuous-process predictions use common assignment rules
for three model inputs. For \(e_{\Gamma}/e_{\max}\), the central value is 1.0;
varying this ratio from 0.9 to 1.1 generates the prediction sensitivity band.
Published sand critical-state and state-parameter data support this range
\cite{woo2019numerical,zhu2020parameter,liu2024evaluation,kurniadi2025drained}.
Poisson's ratio is fixed at \(\nu=0.30\), within reported ranges for granular
soils\cite{suwal2013poisson,thota2021poisson}.

Kim and Kim (2020)\cite{kim2020soil} report case-specific
\(k_i/k_{\mathrm{out}}\) values ranging from 1.26 to 1.46; a representative
value of 1.3 is adopted for those cases. The other prediction cases use
\(k_i/k_{\mathrm{out}}=3.0\), following suction-caisson installation
guidance\cite{houlsby2005design}. The equivalent model ratio \(k_r/k_v\) is
assigned these same values as a model assumption.

The formula reduction and volume closure cases are evaluated independently of
these three inputs.

% ---------------------------------------------------------
% S4.3 Outer-Domain Boundary Rule
% ---------------------------------------------------------
\subsection*{\texorpdfstring{S4.3 Outer-Domain Boundary Rule for \(R_e^*\)}{S4.3 Outer-Domain Boundary Rule for R\_e\^{}*}}
\label{supp:s4-3-1-outer-domain-boundary-rule}

The outer-domain boundary rule determines whether the seepage kernel uses the
unbounded modal estimate \(R_{e,\infty}\) or a finite radius derived from the
experimental boundary. Section S1.3.1 defines the adopted radius \(R_e^*\);
when a source-based radius \(R_b\) is available, it caps
\(R_{e,\infty}\). Because \(R_b\) represents the experimental boundary,
it remains fixed during the sensitivity analysis of
\(e_{\Gamma}/e_{\max}\).

\textbf{Supplementary Table 10. Source-based outer-domain boundary rules.}

\begin{table}[H]
\centering
\scriptsize
\begin{tabularx}{\textwidth}{>{\raggedright\arraybackslash}X
>{\raggedright\arraybackslash}X >{\raggedright\arraybackslash}X
>{\raggedright\arraybackslash}X}
\toprule
Case group & External-boundary basis & Adopted rule & Source class \\
\midrule
Kim and Kim (2020)\cite{kim2020soil} & 900-mm-diameter centrifuge container at 25g and 50g & \(R_b=0.45N_g\) m & \texttt{derived} \\
Tran and Randolph (2008)\cite{tran2008variation} & 390-mm by 650-mm strongbox at 100g & \(R_b=0.195\times100=19.5\) m & \texttt{derived} \\
Tjelta (1995)\cite{tjelta1995geotechnical} & Full-scale field bucket without a test-box boundary & \(R_e^*=R_{e,\infty}\); no finite \(R_b\) & \texttt{direct} \\
Wang et al.~(2019)\cite{wang2019installation} & 1.50-m by 0.96-m by 0.90-m model box & \(R_b=0.48\) m & \texttt{derived} \\
Zhao et al.~(2025)\cite{zhao2025formation} & 1.0-m by 0.5-m by 0.6-m model tank & \(R_b=0.25\) m & \texttt{derived} \\
\bottomrule
\end{tabularx}
\end{table}

Here \(N_g\) is the centrifuge acceleration factor. For each prediction case
with a finite cutoff, \(R_{e,\infty}\) remains below that cutoff; Tjelta has no
finite cutoff. The predictions therefore use \(R_{e,\infty}\) in all evaluated
prediction cases.

% ---------------------------------------------------------
% S4.4 Parameter Sources
% ---------------------------------------------------------
\subsection*{S4.4 Parameter Sources}
\label{supp:s4-4-adopted-benchmark-parameters-and-source-basis}

The parameter-source ledger contains 122 entries. Supplementary Table 11
summarizes model inputs shared across runs, case-assignment rules, and numerical
settings. Supplementary Table 12 shows how data obtained from individual
studies enter each benchmark comparison.

\textbf{Supplementary Table 11. Shared model inputs, assignment rules, and
numerical settings.}

\begin{table}[H]
\centering
\scriptsize
\begin{tabularx}{\textwidth}{>{\raggedright\arraybackslash}X
>{\raggedright\arraybackslash}X >{\raggedright\arraybackslash}X
>{\raggedright\arraybackslash}X}
\toprule
Setting & Adopted value or rule & Source basis & Use \\
\midrule
\(e_{\Gamma}/e_{\max}\) & 1.0 central; 0.9--1.1 band & Declared sensitivity rule & Prediction sensitivity band \\
\(\lambda_c\) & 0.019 & Same-form silica-sand CSL prior & State update \\
\(\kappa_c\) & 0.7 & Same-form silica-sand CSL prior & State update \\
\(p_a\) & 100 kPa & Reference pressure paired with the CSL prior & State update \\
\(\kappa_s\) & 0.006 & Fixed global coefficient & \(M_S\) update \\
\(\nu\) & 0.30 & Granular-soil literature range & Lateral-stress update \\
\(k_r/k_v\) & 1.3 for Kim and Kim; 3.0 otherwise & Model assumption: same value as the adopted \(k_i/k_{\mathrm{out}}\) & Seepage kernel \\
\(p'_{\min}\) & 0.1 kPa & Numerical lower bound & Stress and state update \\
\(\epsilon_{\eta}\) & \(10^{-8}\) & Denominator lower bound & \(M_S\) mobilization \\
\(Q\) & 10 & Bolton-type quartz/silica-sand prior & \(M_D\) correction \\
\(c_{\psi}\) & 0.5 & Fixed Bolton-type coefficient & \(M_D\) correction \\
\(p_{a,B}\) & 1 kPa & Reference-pressure convention & \(M_D\) correction \\
\(\delta_y\) & 0.001 m & Fixed characteristic displacement & Eq.~\eqref{eq:supp-e-6} \\
Material grid & 121 uniformly spaced nodes & Fixed numerical configuration & All solver runs \\
Penetration sequence & Sorted unique benchmark depths & Direct benchmark input & All solver runs \\
Suction sequence & Value paired with each retained depth; no solver interpolation & Direct benchmark input & All solver runs \\
Spatial integration & Trapezoidal rule with linear terminal-interval interpolation & Fixed numerical configuration & Plug-volume assembly \\
Relaxation factor & 0.7 & Fixed numerical configuration & Heave iteration \\
Convergence & Relative tolerance \(10^{-5}\) or residual \(10^{-8}\) m & Eq.~\eqref{eq:supp-d-20} & Heave iteration \\
Iteration limit & 120 per step; reject non-converged steps & Fixed numerical configuration & Run acceptance \\
\bottomrule
\end{tabularx}
\end{table}

\clearpage

\textbf{Supplementary Table 12. Study data and their use in each benchmark
comparison.}

\begin{table}[H]
\centering
\scriptsize
\begin{tabularx}{\textwidth}{>{\raggedright\arraybackslash}X
>{\raggedright\arraybackslash}X >{\raggedright\arraybackslash}X
>{\raggedright\arraybackslash}X}
\toprule
Study & Benchmark comparison & Data obtained from the study & Use in the comparison \\
\midrule
Koteras and Ibsen (2019)\cite{koteras2019medium} & Volume closure, six tests & Geometry, penetration, final heave, and before/after void ratio are direct; area ratio is derived. & Closure uses the reported void-ratio change. \\
Ragni et al.~(2020)\cite{ragni2020observations} & Formula reduction, one test & Prototype geometry, penetration and heave ratios, geometric decomposition, and PIV strain are direct. & The retained test uses the reported self-weight boundary condition. \\
Kim and Kim (2020)\cite{kim2020soil} & Final state prediction, four profiles & Geometry, final heave, suction--depth curves, and most soil properties are direct; missing state inputs follow cited analog or common rules. & Digitized normalized curves are converted at prototype scale; \(k_r/k_v=1.3\). \\
Tran and Randolph (2008)\cite{tran2008variation} & Final state prediction, five profiles & Geometry, suction curves, and most soil properties are direct; initial state is derived where required. & Final heave is calculated as \(h_{\mathrm{heave}}=L-z_{\mathrm{end}}\), where \(L\) is the caisson length and \(z_{\mathrm{end}}\) is the final penetration depth digitized from the published curve. \\
Tjelta (1995)\cite{tjelta1995geotechnical} & Full-scale process prediction, A3 & Field suction and heave are direct; soil-state inputs combine site/design information and cited regional analogs. & The solver uses \(\phi_{\mathrm{model}}=38.5^\circ\); no finite outer boundary is imposed. \\
Wang et al.~(2019)\cite{wang2019installation} & Laboratory process prediction & Geometry, process curves, and the principal soil properties are direct or derived from reported values. & Case inputs use the common hydraulic-ratio and prediction sensitivity band rules in S4.2. \\
Zhao et al.~(2025)\cite{zhao2025formation} & Model-tank process prediction, three tests & Geometry, process curves, relative density, permeability, unit weight, and friction angle are direct; void-ratio bounds use a cited same-group analog. & The initial void ratio is derived from the reported relative density and adopted bounds. \\
\bottomrule
\end{tabularx}
\end{table}
\subsection*{S4.5 Benchmark Data and Summary Metrics}
\label{supp:s4-5-benchmark-data-and-summary-metrics}

% .........................................................
% S4.5.1 Formula Reduction and Volume Closure
% .........................................................
\subsubsection*{S4.5.1 Formula Reduction and Volume Closure}
\label{supp:s4-5-1-formula-reduction-and-volume-closure}

Formula reduction and volume closure assess whether the overall volume relation
reproduces final plug heave from the end-state information reported for each
case. The six Koteras and Ibsen (2019) cases
use the reported void-ratio changes to close the volume balance, whereas Ragni
et al.~(2020) T3 uses the closed-form expression with its reported boundary
condition.

\textbf{Supplementary Table 13. Formula reduction and volume closure summary.}

\begin{table}[H]
\centering
\scriptsize
\begin{tabularx}{\textwidth}{>{\raggedright\arraybackslash}X >{\raggedright\arraybackslash}X >{\raggedright\arraybackslash}X >{\raggedright\arraybackslash}X >{\raggedright\arraybackslash}X >{\raggedright\arraybackslash}X}
\toprule
Case
 & 
Test
 & 
Measured heave (mm)
 & 
Predicted heave (mm)
 & 
Error (\%)
 & 
Source basis
 \\
\midrule
Koteras and Ibsen (2019) & T1 & 45.00 & 49.53 & 10.07 & Reported void-ratio closure. \\
Koteras and Ibsen (2019) & T2 & 47.00 & 48.57 & 3.33 & Reported void-ratio closure. \\
Koteras and Ibsen (2019) & T3 & 44.00 & 47.60 & 8.19 & Reported void-ratio closure. \\
Koteras and Ibsen (2019) & T4 & 42.00 & 54.40 & 29.54 & Reported void-ratio closure. \\
Koteras and Ibsen (2019) & T5 & 49.00 & 55.40 & 13.06 & Reported void-ratio closure. \\
Koteras and Ibsen (2019) & T9 & 57.00 & 58.45 & 2.55 & Reported void-ratio closure. \\
Ragni et al.~(2020) & T3 & 10.50 & 11.19 & 6.60 & Closed-form evaluation with the reported Ragni boundary condition. \\
\bottomrule
\end{tabularx}
\end{table}

Six of the seven calculated heaves differed from the measurements by 13.06\% or
less. The Ragni closed-form case had an error of 6.60\%, whereas Koteras and
Ibsen T4 had the largest error at 29.54\%.

% .........................................................
% S4.5.2 Final State Prediction
% .........................................................
\subsubsection*{S4.5.2 Final State Prediction}
\label{supp:s4-5-2-final-state-prediction}

Final state prediction compares measured final plug heave with the central
prediction and predefined prediction sensitivity band for nine profiles. The four Kim and
Kim endpoints are reported final heaves. The five Tran and Randolph endpoints
are reconstructed from the final digitized penetration depth using
\(h_{\mathrm{heave}}=L-z_{\mathrm{end}}\) under a plug-filled
endpoint assumption. The prediction sensitivity band is obtained by varying
\(e_{\Gamma}/e_{\max}\) over its declared sensitivity range while holding
\(\nu\) at 0.30.

\textbf{Supplementary Table 14. Final state prediction summary.}

\begin{table}[H]
\centering
\scriptsize
\begin{tabularx}{\textwidth}{>{\raggedright\arraybackslash}X >{\raggedright\arraybackslash}X >{\raggedright\arraybackslash}X >{\raggedright\arraybackslash}X >{\raggedright\arraybackslash}X >{\raggedright\arraybackslash}X >{\raggedright\arraybackslash}X}
\toprule
Case
 & 
Profile
 & 
Measured final heave (cm)
 & 
Central final heave (cm)
 & 
Band min (cm)
 & 
Band max (cm)
 & 
Central error (\%)
 \\
\midrule
Kim and Kim (2020) & T1-1 S 25g & 66.00 & 48.26 & 37.00 & 58.31 & -26.87 \\
Kim and Kim (2020) & T1-2 S 50g & 53.00 & 52.30 & 39.47 & 64.53 & -1.32 \\
Kim and Kim (2020) & T3-1 S 50g & 58.00 & 50.74 & 38.28 & 63.13 & -12.52 \\
Kim and Kim (2020) & T4-3 S 25g & 70.00 & 59.47 & 46.04 & 73.80 & -15.05 \\
Tran and Randolph (2008) & TRAN-T1 & 50.85 & 60.53 & 48.07 & 70.46 & 19.04 \\
Tran and Randolph (2008) & TRAN-T2 & 45.04 & 41.63 & 34.20 & 47.79 & -7.56 \\
Tran and Randolph (2008) & TRAN-T3 & 45.04 & 32.37 & 27.08 & 37.34 & -28.13 \\
Tran and Randolph (2008) & TRAN-T4 & 59.93 & 59.67 & 47.34 & 72.31 & -0.43 \\
Tran and Randolph (2008) & TRAN-T5 & 55.75 & 55.42 & 43.50 & 63.53 & -0.58 \\
\bottomrule
\end{tabularx}
\end{table}

Seven of the nine measured final heaves fell within the prediction sensitivity bands. The
two outside-band profiles, Kim and Kim T1-1 S 25g and Tran and Randolph
TRAN-T3, had central errors of \(-26.87\%\) and \(-28.13\%\), respectively.

% .........................................................
% S4.5.3 Continuous Process Prediction
% .........................................................
\subsubsection*{S4.5.3 Continuous Process Prediction}
\label{supp:s4-5-3-continuous-process-prediction}

Continuous process prediction compares final plug heave and the measured
heave--depth curve for five profiles. For all five profiles, final-point error
is evaluated at the last measured depth. The field curves reported by Tjelta
overlap substantially. The A3 trace was the clearest and most continuous over
the installation depth and was therefore digitized for the process comparison.

The complete curve is summarized by curve mean absolute percentage error
(MAPE), root-mean-square error (RMSE), and normalized RMSE (NRMSE). Curve NRMSE
is the curve RMSE normalized by the measured final heave. Because pointwise
percentage errors can become large when measured heave is small during early
penetration, curve MAPE is interpreted alongside final-point error, absolute
RMSE, and NRMSE.

\textbf{Supplementary Table 15. Final-point comparison for continuous process
prediction.}

\begin{table}[H]
\centering
\scriptsize
\begin{tabularx}{\textwidth}{>{\raggedright\arraybackslash}X
>{\raggedright\arraybackslash}X >{\raggedright\arraybackslash}X
>{\raggedright\arraybackslash}X >{\raggedright\arraybackslash}X
>{\raggedright\arraybackslash}X >{\raggedright\arraybackslash}X}
\toprule
Case
 & 
Profile
 & 
Measured final point (cm)
 & 
Central final point (cm)
 & 
Band min (cm)
 & 
Band max (cm)
 & 
Final-point error (\%)
 \\
\midrule
Tjelta (1995) & TJELTA-A3 & 15.43 & 16.70 & 14.45 & 18.98 & 8.23 \\
Wang et al.~(2019) & full process & 4.79 & 4.96 & 4.51 & 5.41 & 3.45 \\
Zhao et al.~(2025) & ZHAO-T1 & 3.68 & 3.73 & 3.40 & 4.06 & 1.29 \\
Zhao et al.~(2025) & ZHAO-T2 & 3.48 & 3.67 & 3.37 & 3.98 & 5.49 \\
Zhao et al.~(2025) & ZHAO-T3 & 3.28 & 3.37 & 3.10 & 3.63 & 2.74 \\
\bottomrule
\end{tabularx}
\end{table}

\textbf{Supplementary Table 16. Curve-error metrics for continuous process
prediction.}

\begin{table}[H]
\centering
\scriptsize
\begin{tabularx}{\textwidth}{>{\raggedright\arraybackslash}X
>{\raggedright\arraybackslash}X >{\raggedright\arraybackslash}X
>{\raggedright\arraybackslash}X >{\raggedright\arraybackslash}X}
\toprule
Case & Profile & Curve MAPE (\%) & Curve RMSE (cm) & Curve NRMSE (\%) \\
\midrule
Tjelta (1995) & TJELTA-A3 & 95.49 & 3.15 & 20.42 \\
Wang et al.~(2019) & full process & 19.24 & 0.52 & 10.80 \\
Zhao et al.~(2025) & ZHAO-T1 & 36.73 & 0.16 & 4.22 \\
Zhao et al.~(2025) & ZHAO-T2 & 119.25 & 0.14 & 3.92 \\
Zhao et al.~(2025) & ZHAO-T3 & 13.88 & 0.14 & 4.25 \\
\bottomrule
\end{tabularx}
\end{table}

All five measured final points fell within the prediction sensitivity bands, with
final-point errors ranging from 1.29\% to 8.23\%. Across the complete curves,
NRMSE ranged from 3.92\% to 4.25\% for the Zhao profiles, compared with
10.80\% for Wang and 20.42\% for Tjelta.

% ---------------------------------------------------------
% Supplementary References
% ---------------------------------------------------------
\renewcommand{\refname}{Supplementary References}
\phantomsection
\label{supp:supplementary-references}

%\pagenumbering{Roman}

% \input{response_letter}

\end{document}